\documentclass[preprint2]{aastex}

\newcommand{\etal}{{et al.~}}
\newcommand{\kms}{km s$^{-1}$}
\newcommand{\meanalpha}{\langle\alpha\rangle}
\newcommand{\meanBfield}{\langle B\rangle}
\newcommand{\be}{\begin{equation}}
\newcommand{\ee}{\end{equation}}
\newcommand{\bperp}{$B_\perp$}
\newcommand{\blos}{$B_{los}$}
\newcommand{\hii}{\ion{H}{2}}
\newcommand{\intfil}{Integral-shaped Filament}

\shorttitle{Polarimetry of Cores in Orion B}
\shortauthors{Matthews, Fiege \& Moriarty-Schieven}

\begin{document}

\title{Magnetic Fields in Star-Forming Molecular Clouds III. Submillimeter Polarimetry of Intermediate Mass Cores and Filaments in Orion B}

\author{Brenda C. Matthews} 

\affil{Department of Physics and Astronomy, McMaster University,
Hamilton, ON, Canada L8S 4M1}

\email{matthews@physics.mcmaster.ca}

\altaffilmark{1}
\altaffiltext{1}{Current address: Dept. of Astronomy, UC, Berkeley (bmatthews@astro.berkeley.edu)}

\author{Jason D. Fiege}

\affil{Canadian Institute for Theoretical Astrophysics, University
of Toronto, Toronto, ON, Canada M5S 3H3}

\email{fiege@cita.utoronto.ca}

\and

\author{Gerald Moriarty-Schieven}

\affil{National Research Council of Canada, Joint Astronomy Centre,
Hilo, HI U.S.A. 96720} 

\email{g.moriarty-schieven@jach.hawaii.edu}

\begin{abstract}

Using the imaging polarimeter for the Submillimeter Common User
Bolometric Array at the James Clerk Maxwell Telescope, we have
detected polarized thermal emission at 850 \micron\ from dust toward
three star-forming core systems in the Orion B molecular cloud: NGC
2071, NGC 2024 and LBS 23N (HH 24).  The polarization patterns are not
indicative of those expected for magnetic fields dominated by a single
field direction, and all exhibit diminished polarization percentages
toward the highest intensity peaks.  NGC 2024 has the most organized
polarization pattern which is centered consistently along the length
of a chain of 7 far-infrared sources.  We have modeled NGC 2024 using
a helical field geometry threading a curved filament and also as a
magnetic field swept up by the ionization front of the expanding \hii\
region.  In the latter case, the field is bent by the dense ridge,
which accounts for both the polarization pattern and existing
measurements of the line-of-sight field strength toward the northern
cores FIR 1 to FIR 4.  The direction of the net magnetic field
direction within NGC 2071 is perpendicular to the dominant outflow in
that region. Despite evidence that line contamination exists in the
850 \micron\ continuum, the levels of polarization measured indicate
that the polarized emission is dominated by dust.

\end{abstract}


\section{Introduction}
\label{p3:intro}

The Orion B (L1630) molecular cloud, at a distance of 415 pc
\citep{ant82}, is one of the nearest giant molecular clouds and is an
active site of low- to high-mass star formation.  It was one of the
first clouds to be systematically studied for dense cores by
\citet{lbs91}, who found that massive star formation takes place only
in the five largest clumps, which together make up more than 50\% of
the mass of dense gas.  We have chosen three of these five clumps, NGC
2071IR (LBS 8), NGC 2024 (LBS 33), and HH24 (LBS 23), for the current
study.  A fourth region, NGC 2068, contains a string of substantially
smaller, fainter cores connected by weak dusty filaments
\citep{mit01}.  Polarimetry of this region will be presented in
a forthcoming paper \citep[Paper IV in this series]{mw01}.

Although the three regions have comparable gas masses, ranging from
230-460 M$_\odot$ \citep{lbs91}, they have very different star
formation properties.  NGC 2071IR lies four arcminutes north of the
reflection nebula NGC 2071.  This extended submillimeter source
consists of a cluster of at least nine embedded infrared stars
\citep{wal93} with a combined infrared luminosity of 520 L$_\odot$
\citep{but90}.  The source IRS 3 is thought to be the driving source
of a massive bipolar molecular outflow
\citep{bal82,sne84,mor89,eis00}.  \citet{hou00} infer alignment
between the outflow and its magnetic field by comparison of spectral
lines of neutral and ionic species.  Shocked molecular hydrogen
\citep{bla82} and H$_2$O masers \citep{gen79} are also seen towards
this region, which is in a later evolutionary stage than NGC 2024
\citep{lau96}.  \citet{eis00} has also documented several other
outflows in the region.  Submillimeter maps of this region have been
published by several authors \citep{mit01,mot01,joh01}.  Based on a
comparison between submillimeter continuum and CO(3-2) and HCO$^+$
line data, \citet{mot01} suggest that $> 20$\% of the 850 \micron\
emission in the outflow region could originate from line
contamination.

By contrast with the other two regions, the HH 24-26 (LBS 23) cores
have relatively little extended submillimeter emission and are mostly
compact \citep{mit01,joh01,lau96,lis99} and cold ($<$10K,
\citet{chi93}).  Of the twelve condensations identified by
\citet{lis99}, our polarimetric image covers LMZ 2, 3, and 4.  All
three of these cores have 3.6 cm continuum sources
\citep{baw95,gib99}.  LMZ 3 (also known as HH24MMS, \citet{chi93}) is
a class 0 protostar \citep{baw95}, while LMZ 4 is a T Tauri star with
a known CO outflow \citep{sne82,gib93}.  These are all indicators of
low-mass star formation.

NGC 2024 has an associated \hii\ region and is the most prominent star
formation region in Orion B, associated with a massive cluster,
ionizing B stars, and stars at all phases of evolution
\citep{mez88,lad91,cc96}.  The submillimeter continuum emission was
discovered by \citet{mez88}.  The emission arises from a dense ridge
of gas and dust behind the \hii\ region, determined from the velocity
of associated gas \citep{cru86,bar89}, and consists of at least seven
sources aligned along a ridge, similar to OMC-3 in Orion A
\citep{joh99,chi97}.  Relatively more mass exists in the filamentary
gas in the latter region.  Two of these cores (FIR 4 and FIR 5) are
the origins of unipolar molecular outflows, one of which is very
highly collimated and very extended \citep{san85,ric92,cc96}, while
the FIR 6 core exhibits a compact outflow \citep{cc96} and contains a
water maser \citep{gen77}, a signature of intermediate-mass
protostars.  The rest of the cores show no sign of star formation
activity \citep{vis98}.  \citet{fuk00} present a numerical simulation
of triggered star formation along a filament by the expansion of an
\hii\ region and apply their analysis to NGC 2024 in particular.  In
their picture, compression triggers two cores (i.e., FIR 4 and FIR 5)
to collapse, and then at a later time, further collapse is triggered
further up and down the filament.  This sequence was observed by
\citet{cc96} in their study of outflows from NGC 2024, where the
dynamical ages of outflows from FIR 5, FIR 4 and FIR 6 are $1.4 \times
10^4$ yr, $2.6 \times 10^3$ yr, and 400-3300 yr.

Magnetic fields play a crucial role in the process of star formation,
through the magnetic support of molecular clouds, dissipation of
angular momentum in accretion disks, and the generation of jets and
outflows (see \citet{hei93} and references therein).  Polarized
thermal emission at submillimeter wavelengths from aligned dust grains
traces the direction of the magnetic field structure projected onto
the plane of the sky \citep{hil88}.  Absorption and scattering are
usually negligible in the submillimeter.  This is in contrast to
optical and near infra-red polarimetry, where both of these processes
can cause contamination.  With the recent development of focal plane
bolometer arrays equipped with polarimeters, sensitive imaging
polarimetry in the submillimeter is now possible.

Polarimeters functioning at 100 \micron\ (aboard the Kuiper Airborne
Observatory), 350 \micron\ (at the Caltech Submillimeter Observatory),
and 850 \micron\ (at the James Clerk Maxwell Telescope) have detected
polarized emission from dust toward many Galactic molecular clouds.
These include the well-studied OMC-1 core in Orion
\citep{sch98,sch97,cop00,dot00}; Sagittarius A \citep{ait00};
Sagittarius B2 \citep{dow98} and recently OMC-3 in Orion
(\citet[hereafter Paper I]{mw00}; \citet[hereafter Paper II]{mwf01};
Dowell 2001, in preparation).  Numerous protostellar and starless
cores have also been mapped (i.e., \citet{hol96,war00}).

Polarimetry provides information only on the plane-of-sky magnetic
field orientation, but no direct information about the magnetic field
strength, since the degree of polarization is also dependent on other
factors such as grain shape, degree of alignment, and composition.  To
obtain information about the magnetic field strength requires
observation of Zeeman splitting of molecular or atomic spectral lines,
which additionally provides information about the direction of the
field along the line of sight.  

In this paper, we present the first submillimeter polarimetry of the
NGC 2071 core and the LBS 23N region.  Zeeman observations do not
exist toward either of these two clouds.  Far-infrared polarimetry at
100 \micron\ \citep{dot00} and Zeeman splitting observations of OH
\citep{cru83,kaz86,crtg99} exist for the NGC 2024 ridge of cores.  In $\S$
\ref{p3:obs}, we describe the 850 \micron\ observations and the data
reduction.  The data are presented in $\S$ \ref{p3:sec:results} and the
polarization patterns are interpreted in $\S$ \ref{p3:sec:theory}.  We
summarize our results in $\S$ \ref{p3:summary}.

\section{Observations and Data Reduction}
\label{p3:obs}

The observations presented here were obtained on 1998 September 7 and
8 using SCUBA (Submillimeter Common User Bolometer Array)
\citep{hol99} at the James Clerk Maxwell Telescope\footnote{The JCMT
is operated by the Royal Observatory Edinburgh on behalf of the
Particle Physics and Astronomy Research Council of the United Kingdom,
the Netherlands Organization for Scientific Research, and the National
Research Council of Canada.}  combined with the SCUBA polarimeter
\citep{gre00,gre01b}.  The nights were very stable, with
$\tau$(225GHz) ranging from 0.05 to 0.07 during the period of
observations.  Calibration of the polarizer was performed on 1998
September 5 using the Crab Nebula, for which a percentage polarization
$p = 19.3 \pm 4\%$ and position angle $\theta = 155 \pm 5^{\circ}$
were measured.

Observations were made of two overlapping fields toward NGC 2024,
and one field each toward NGC 2071 and LBS 23N.  Each polarization cycle
consists of 16 integrations at 22.5$^\circ$ rotation intervals 
in 8 minutes integration time.  Six observations (48 minutes
integration time) were made toward NGC 2071 and each position toward
NGC 2024, and 27 observations (3.6 hours integration time) were made
toward LBS 23N.

For each polarization cycle map, the raw SCUBA data were reduced using
standard SCUBA software \citep{hol99} to perform nod compensation,
flatfielding and extinction corrections.  More information on the
polarimeter and its data cycle can be found in \citet{gre00}.  For
polarimetry data, extinction is estimated by extrapolation from the
current CSO tau value at 225 GHz.  Bolometers with anomalously high
noise were flagged, and the data were clipped at the 10 $\sigma$
level.  Sky noise and instrumental polarization (IP) removal was then
performed.  The removal of sky noise is, even for basic SCUBA data, a
task fraught with danger, but is even more so for polarimetry.  The
techniques of sky removal are discussed in detail in Paper II; we
have taken care to select empty bolometers for sky subtraction.

To first order, the removal of sky effects is achieved by chopping the
JCMT secondary during SCUBA observations.  Systematic errors can be
introduced if significant polarized flux exists at the chop, or
reference, positions.  Large chop throws of 150\arcsec\ were used for
these observations (180\arcsec\ is the maximum for the JCMT) to try to
chop as far off the bright emission as possible.  The possible effects
of polarized flux in the reference beams (or bolometers used for sky
removal) are discussed in Paper II.

Since we have access to large-scale 850 \micron\ scan maps of Orion B
north \citep{mit01} and south obtained by the Canadian Consortium for
Star Formation Studies at the JCMT, comparisons can be made between
intensities of candidate bolometers for sky noise subtraction in the
on-source position and the fluxes in the corresponding chop positions.
For two of our regions, NGC 2024 and NGC 2071, comparison reveals that
the differences between these two positions are significant, and that
no bolometers on-source were ``empty''.  The region around LBS 23N, on
the other hand, reveals the same flux levels at the chop positions as
at the edges of the on-source pointing.  Thus, some bolometers at the
edges of that map had average intensities of zero.  To deal with sky
bolometers which do not contain zero flux, the mean value per
bolometer removed during sky noise subtraction is added back into the
maps of NGC 2024 and NGC 2071, increasing the intensity in each
bolometer.  The main consequence of this addition is to reduce the
polarization percentage, which is inversely proportional to $I$.

After correcting for source rotation across the array, the Stokes'
parameters were calculated by comparing measurements offset by
45$^\circ$ in waveplate rotation (90$^\circ$ on the sky).  The $I$,
$Q$, and $U$ maps for each pointing center were averaged, and standard
deviations were derived by comparing the individual data sets.  The
two overlapping fields toward NGC 2024 were then combined into a
mosaic; polarization data are uncalibrated, so the data were
combined simply by averaging.  Since no absolute flux calibrations are
done on polarimetry data, the zero points are unknown.  This presents
problems for creating maps requiring more than one SCUBA field, since
one does not know which field contains the best base level to use as
reference.  We want to obtain the best estimates of uncalibrated flux
without giving undue weight to either of our mapped fields.  This was
done by averaging the baseline levels in the two maps.

The maps were then binned spatially by a factor of 3 (NGC 2071) or 4
(NGC 2024 and LBS 23N) in both RA and DEC to yield 9 or
12\arcsec~sampling.  Selection of ``good'' polarization vectors is
done by filtering out all vectors for which $p<1$\% (since the IP
values are accurate to just $\pm 0.5$\%) and then filtering by
signal-to-noise of polarization percentage, uncertainty of
polarization percentage, and the intensity level as appropriate to
each field.  The thresholding of $p$ ensures that the effects of
sidelobe polarization are low (see $\S$ \ref{p3:sidelobe}).  As further
confirmation of the observed polarization patterns, we divided the
data sets for each source into two subsets and performed complete
reduction on each.  The polarization patterns were qualitatively
consistent with one another in all cases.

We have truncated our selection of vectors at levels of 2\% of the NGC
2071IR peak intensity, 2.5\% of the NGC 2024 source peak and 4\% of
the LBS 23N peak.  Paper II illustrates that percentage
polarizations and position angles are reliable down to these levels in
cases where the reference position's intensity is approximately 2\% of
the source peak intensity and the polarization percentage in the
reference position is no greater than that of the source field.  Thus,
in the lowest flux regions, the polarization percentage could be
overestimated by up to a factor of two while the polarization position
angle is incorrect by $< \pm 10^\circ$.  Since JCMT data obtained by
chopping off the source record only the difference between these two
positions, our data provide no information about the polarized
emission at the reference position.  However, based on more extensive
intensity maps produced by the scan-mapping technique, the chop
positions are approximately 1\%, 2\% and $<1$\% of the on-source peaks
for NGC 2071IR, NGC 2024, and LBS 23N respectively.  These are
effectively lower limits to the fractional fluxes relative to the
peaks, since scan mapping data may remove the fluxes of very extended
features.  If there is a uniform background chopped out of the scan
maps as well, then the reference position could in fact contain a more
significant fraction of the source peak flux.

\subsection{Sidelobe Polarization}
\label{p3:sidelobe}

Even though the sidelobes of the JCMT beam contain less than 1\% of
the main beam power at 850 \micron, significant polarization can be
measured there because the two incoming planes of polarization
experience different optical effects.  This can produce significant
{\it sidelobe polarization}, that can become a source of systematic
error especially when mapping extended fields in which the source
located in the main beam is fainter than sources elsewhere in the
SCUBA field.  One can estimate the minimum believable source
polarization percentage from the expression:
\begin{equation}
p_{crit} \ge 2 \times p_{sl} \left(\frac{P_{sl}}{P_{mb}}\right)
\left(\frac{F_{sl}}{F}\right)
\label{p3:pcrit}
\end{equation}

\noindent where $p_{sl}$ is the IP in the relevant part of the
sidelobe, $P_{sl}/P_{mb}$ is the ratio of the sidelobe power to that
in the main beam, and $F_{sl}/F$ is the ratio of the mapped source
flux in the sidelobe to that in the main beam \citep{gre01b}.
Basically, $p_{crit}$ is an estimate of the polarization percentage
produced by the sidelobe source alone; hence only polarizations in
excess of that value are believable.

For the regions of NGC 2071 and LBS 23N, sidelobe polarization will be
minimal, since in both those fields, the field centers were located at
the submillimeter peaks.  However, in NGC 2024, our northern mapping
field was centered roughly on FIR 3, but FIR 5 was at the edge of the
SCUBA map, approximately 70\arcsec\ from the center.  The ratio of the
FIR 5 flux to the central flux is $\sim 3.4$.  Using a map of Saturn
from 31 August 1998, the average power at 70\arcsec\ from the field
center is approximately 0.3\% of the main beam.  The average
instrumental polarization is 5.5\%.  Substitution in equation
(\ref{p3:pcrit}) yields $p_{crit} \ge 0.11$\%.  The planetary map was
generated with a chop throw of 120\arcsec, somewhat less than the
150\arcsec\ chop used in NGC 2024.  However, using planetary data on
Saturn from 11 October 1999 with a 150\arcsec\ chop, we obtain an
estimate of $p_{crit} \ge 0.2$\%.  Hence, we are confident that our
thresholding of polarization percentage at 1\% removes all effects of
sidelobe polarization in our maps.

\section{The Polarization Data}
\label{p3:sec:results}

The polarized emission in the submillimeter is thermal emission
originating from rapidly spinning non-spherical grains, which are
expected to be aligned, on average, perpendicular to the magnetic
field.  The precise details of the alignment processes are not
entirely understood.  However, it is well understood that
supra-thermally rotating grains align perpendicular to the magnetic
field on very short time-scales.  Thus, grain alignment depends
critically on the mechanisms by which grains are accelerated to
supra-thermal rotational velocities.  Radiative torques from short
wavelength radiation \citep{dra96} and the well-known Purcell
``rocket'' mechanism \citep{pur79} have both been proposed as
candidate mechanisms.  At present, however, these mechanisms appear
insufficient to accelerate grains to the required rotational speeds in
regions of high optical depth within molecular clouds \citep{laz97}.
However, the data presented in this paper, as well as recent papers by
other authors (Paper II; \citet{dot00,sch98}) provide evidence that grain
alignment does occur in dense regions of molecular clouds and that the
underlying field structure is well-ordered, as demonstrated by the
highly structured polarized emission seen in our maps.  \citet{joh01}
compare 850 \micron\ data of NGC 2071 and LBS 23 to CS $J=2-1$ maps by
\citet{lbs91} and conclude that these clumps of dust emission occur
within regions above a column density threshold of $N(H) \sim 10^{22}$
cm$^{-2}$, which corresponds to an $A_V >5$.  Thus, the 850 \micron\
dust emission is associated with extinctions exceeding those at which
radiative alignment is expected to be significant.  This result poses
an important challenge for grain alignment models.

The shapes of grains, their size distribution, their composition, and
the details of their alignment mechanisms all strongly affect the
ability of grains to produce polarized thermal radiation, and hence
influence the resulting polarization patterns.  None of these factors
are well-constrained, either theoretically or observationally.
However, \citet{fp00c} developed a pragmatic approach that combines
{\em all} of the grain and alignment properties for a distribution of
grain species into a single parameter $\meanalpha$, which can be
obtained observationally.  This parameterization permits modeling of
polarization patterns without detailed knowledge of grain and
alignment physics.  It was assumed in \citet{fp00c} that all grain
species are uniformly mixed with the gas and aligned to the same
extent everywhere within the cloud, which implies that $\meanalpha$ is
constant within any given cloud.  However, their model is easily
extended to allow systematic variations in grain and alignment
properties throughout the cloud, by simply allowing $\meanalpha$ to be
a function of either the density, or the distance from the centre of
the cloud.  We find that allowing $\meanalpha$ to vary within a cloud
does not typically have a large impact on the geometry of the
polarization patterns, although the polarization percentage and the
degree of depolarization toward bright sources can vary more
substantially.  In this sense, our models are quite robust against
changes to the underlying grain composition and alignment properties.

If we assume, for now, that $\meanalpha$ is constant, then the maximum
possible polarization percentage (for the most favourable geometry)
with a constant field parallel to the plane of the sky, is given by
the formula 
\be p_{max}=\frac{\meanalpha}{1-\meanalpha/6}
\label{p3:eq:pmax}
\ee \citep{fp00c}.  The maximum polarization percentages for the three
regions that we have mapped are 15\% for NGC 2071, 13\% for NGC 2024,
and 22\% for LBS 23N.  The value for LBS 23N may be anomalously high;
the next highest percentage measured is 12.6\%.  Therefore, 15\% is a
reasonable maximum value for NGC 2071, with 13\% an upper limit for
NGC 2024 and LBS 23N.  Using 15\% as an upper limit on the
polarization percentages in these regions, and assuming this is close
to the theoretical maximum for a line of sight passing through a
magnetic field with ideal geometry, then we find that $\meanalpha
\approx 0.14$ for all three regions.  This estimate is really a lower
bound for $\meanalpha$ because the observed polarization percentage is
reduced by any component of the field lying along the line of sight.
Nevertheless, we are encouraged that our estimates of $\meanalpha$ do
not vary enormously between the three regions, and conclude that
$\alpha=0.14$ is a reasonable global estimate for $\meanalpha$ in
Orion B.  We use this value of $\meanalpha$ to generate models of NGC 2024
(see \S \ref{p3:sec:NGC2024}).

\subsection{Polarization Patterns}

\subsubsection{NGC 2071}

In Figure \ref{p3:n2071}, we present the 850 \micron\ polarization
vectors of NGC 2071 IR, superimposed on a false color image of the 850
\micron\ dust emission (from \citet{mit01}).  The inset shows a
``blow-up'' of the central region in order to show the weak
polarization features better.  The data are also presented 
in Table \ref{p3:n2071table} of Appendix \ref{p3:appendixA}.

Toward the region of strongest emission (the IR cluster), the
percentage polarization is quite weak, typically less than 2\%.
However, polarization as strong as 15\% is measured toward fainter
parts of this region.  The polarization pattern is highly symmetric,
including the curvature of the vector orientation around the peak to
the northeast and southwest.  Over the entire core region, the vectors
exhibit a mean orientation of 20$^\circ$ east of north.  Based on
low-flow velocities of HCO$^+$ maps of the dominant outflow in the
region, \citet{gir99b} estimate its overall position angle to be
$\approx 40^\circ$ (east of north).  Figure \ref{p3:n2071} shows this
outflow orientation schematically.  Closer to the emission peak, the
dominant position angle is $\sim 30^\circ$, which is within $10^\circ$
of the outflow's orientation.  The outflow is thus well-aligned with
the polarization vectors.

Qualitatively, the polarization pattern is similar to that of OMC-1,
which was interpreted as evidence for an hourglass magnetic field
geometry in that region \citep{sch98}.  In this picture, the magnetic
field is pinched toward the central source due to flux freezing with
the infalling gas.  However, we do not observe the same sort of
flattening of the NGC 2071 core that is evident in the 350 \micron\
flux map of OMC-1 \citep{sch98}.  The NGC 2071 core maintains a
circular shape even far from the peak, with no suggestion of
oblateness or prolateness (see Figure \ref{p3:n2071}).

\subsubsection{LBS 23N}

Figure \ref{p3:lbs23} shows the dust polarization of the LBS 23N
source and its surrounding region (the data are also presented in
Table \ref{p3:lbs23ntable} of Appendix \ref{p3:appendixA}).  A string
of cores is surrounded by a faint filamentary background of dust
emission.  The dominant orientation of the polarization pattern is
north-south, except at the southern boundary, where the vectors lie
almost east-west.  The strongest polarization (22\%) is seen in the
east and west limbs of the ``filament''.

It is intriguing that the vectors abruptly change direction by
$90^\circ$ near the southern-most boundary of our polarization map to
become aligned east-west.  The 850 \micron\ continuum map of
\citet{mit01} shows that the underlying dust structure widens south of
our polarization map to a filamentary segment oriented roughly
east-west (see Figure \ref{p3:lbs23}).  A similar shift of
polarization orientation has been observed in the southern part of the
OMC-3 filament in Orion A, where the filament's projected width on the
plane of the sky appears to increase (Papers I and II).  We discuss
three possible interpretations of our data in $\S$
\ref{p3:sec:LBS23N}.  Polarimetry in the southern region of LBS 23
will help us in our modeling effort to understand the structure of the
magnetic field in this region.  However, the faintness of the LBS 23
filament will make obtaining the necessary data more difficult than in
OMC-3.

\subsubsection{NGC 2024}

The polarization pattern of the molecular ridge of NGC 2024 is shown
in Figure \ref{p3:n2024}, superimposed on a false color image of the
850 \micron\ dust emission.\footnote{JCMT data taken by Canadian
Consortium for Star-Formation Studies, as yet unpublished.}  The data
are presented in Table \ref{p3:n2024table} of Appendix
\ref{p3:appendixA}.  Seven continuum sources (FIR 1-7) have been
identified along the ridge \citep{mez88,mez92}.  In the north (FIR
1-4), the vectors along the peaks are weak and parallel to the ridge.
The southern sources (FIR 5-7) have very weak polarization toward the
brightest cores and are significantly depolarized.  To the east and
west, the polarization becomes quite intense (up to 10\%), and a
change in vector orientation of approximately 90$^\circ$ is observed
from west to east.

Linear polarization at 100 \micron\ has been observed previously
toward this region with the Kuiper Airborne Observatory
\citep{hil95,dot00}.  These data are centered on the FIR 5 core and
cover the same area as our map.  Figure \ref{p3:2pols} shows our
polarization data and those of \citet{dot00} at 100 \micron\ plotted
over contours of the 850 \micron\ total emission.  The 100 \micron\
data are more sparsely sampled than the SCUBA data (owing to the
35\arcsec\ beam), but the two patterns are remarkably consistent.  The
vectors trace the same V-shaped pattern seen at 850 \micron, although
the finer sampling by SCUBA makes the pattern much easier to see.  The
chop throw of the KAO observations was 10.5\arcmin, over four times
greater than that used for our data set.  The fact that the
polarization patterns are very consistent suggests that the position
angle at least has not been strongly affected by any systematic
effects due to chopping.  The maximum polarization percentage measured
at 100 \micron\ is $5.86^{+0.01}_{-1.00}$\%, while the minimum vector
has a polarization percentage of $0.36^{+0.06}_{-0.07}$\% and a
position angle of $164.9 \pm 5.2^\circ$ (east of north), where the
uncertainties are absolute upper limits.  At roughly the same
position, the 850 \micron\ polarization is $1.63 \pm 0.15$\% at an
angle of $-20.7 \pm 2.7^\circ$ (equivalent to $159.3 \pm 2.7^\circ$).
This is a relatively low polarization percentage at 850 \micron, and
the position angles of the vectors agree within measurement
uncertainties.  At 850 \micron, the maximum and minimum polarization
percentages are $13.41\pm 1.13$\% and $1.02 \pm 0.09$\%.  Where the
northern and southern parts of the filaments meet (i.e., between FIR 4
and 5), there is significant depolarization regardless of the low flux
level. Depolarization in the filamentary material away from the cores
is also observed between cores in OMC-3 in Orion A (see Paper II).

There is no obvious correspondence between the polarization position
angles and the directions of outflows in the region.  The strongest
outflow is unipolar and extends from FIR 5 to the southeast (Richer,
Hills \& Padman 1992).  Greaves, Holland \& Ward-Thompson (2001) have
measured polarized spectral line emission from the CO $J=2-1$ line
utilizing the Goldreich-Kylafis effect \citep{gold81} in the outflow
of FIR 5.  The polarization vectors measured in the outflow are
plotted on Figure \ref{p3:n2024}, along with the orientation of the
outflow itself.  For large CO optical depths, line polarization is
parallel to the field, unless the field is at a large angle to the
velocity gradient \citep{kyl83}.  Therefore, \citet{gre01a} conclude
that the magnetic field associated with the outflowing gas is aligned
with the outflow itself.  If the large scale field in the region
maintains the same orientations, we would expect the continuum
polarization vectors to lie perpendicular to the outflow.  But the
continuum vectors do not exhibit this behavior.  Nor does their
orientation appear to be influenced by their proximity to the outflow.
The dust polarization data do not exhibit an obvious correlation with
either the outflow or the line polarization data which traces its
magnetic field.  Thus, we conclude that a complex field geometry
threads this region.

\subsection{Depolarization toward High Intensities}
\label{p3:sec:depol}

As observed in many other regions, a declining polarization percentage
is detected toward regions of increasing intensity in each of our
fields.  Figure \ref{p3:depol} shows the percentage polarization as a
function of unpolarized $I$ for each region.  The depolarization
effect is clearly a global feature in our maps and is not limited to a
single region or core.  The slopes of Figure \ref{p3:depol} range from
$-0.5$ to $-0.9$, but the trend of declining polarization percentage
is clear.  The same trend was measured in the OMC-3 region of Orion A
(see Paper II).  It is possible for a ``polarization hole'' to result
from a systematic error produced by chopping to remove background sky
during observing.  However, Paper II shows that depolarizations as
steep as those measured in Orion B (i.e., slopes of log $I$ versus log
$p$ $< -0.55$) cannot be produced by the effects of chopping alone,
particularly at high values of $I$, even for the case where the flux
at the reference position is 25\% that of the source peak and the
reference polarization percentage is twice that of the source.
Therefore, there must be a physical cause for this effect.

The issue of depolarization in dense regions is particularly
interesting as a test of alignment mechanisms.  Grain alignment theories
at present cannot explain why grains are so well aligned in
regions of high optical depth within molecular clouds.  For example,
the Purcell mechanism \citep{pur79} and its variations are not likely
to work in these regions, where atomic hydrogen is almost
completely absent \citep{laz97}.  The radiative alignment mechanism
\citep{dra96}, which aligns grains efficiently in the ISM, is unlikely
to work in dense molecular regions, where short wavelength starlight
(needed to spin up and align grains) is extinguished.  Appealing to
embedded sources to provide the necessary radiation is not likely to
help, since this would predict the best alignment to be associated
with the dense regions around evolved protostars, while the opposite
is observed.  Thus, our observations pose an interesting and extremely
important challenge for grain alignment theory.

We consider 3 possible explanations for the ``polarization hole''
effect.  First of all, it is possible that the depolarization in our
maps are the result of grains in the central regions of highest
density that are either poorly aligned, or intrinsically poor
polarizers \citep{wga00,goo95}.  In such a case, all dust grains
through the cloud contribute to the continuum intensity, but only the
fraction that are aligned contribute to the polarized flux given by
the Stokes' vectors $Q$ and $U$.  Therefore, toward regions of high
intensity, the ratio of polarized to total intensities is
systematically reduced.  Alternatively, \citet{fp00c} proposed that
the depolarization could result from the large scale field structure
itself, since helically twisted fields result in depolarization as a
result of oppositely signed contributions to the Stokes' vectors from
the poloidal and toroidal field components.  This field geometry has
been used to model the OMC-3 part of Orion A's \intfil\ (Paper II),
and we show a similar model for NGC 2024 in $\S$ \ref{p3:sec:helical}
below.  A third possibility is that a disordered, or tangled, field
exists on scales smaller than our beam, and the field averages to zero
in the JCMT's 14\arcsec\ beam.  It is possible that more than one of
these effects could be contributing to the depolarization observed in
our maps.

Interestingly, NGC 2071 does not exhibit a polarization hole
coincident with the peak of the dust emission; instead, two
polarization holes are detected, offset to the northeast and southwest
of the dust emission peak.  These polarization holes are not due to
bad bolometers or low signal-to-noise.  The degree of polarization in
those locations is less than 1\% and is thus considered negligible at
these positions.  In NGC 2024, depolarization is so significant in the
case of FIR 5-7 that polarization is negligible across those cores.
In LBS 23N, depolarization can be seen in Figure \ref{p3:lbs23} toward
the brightest core, OriBN 59 (see \citet{mit01}).  It is difficult to
sample the depolarization in the fainter sources of Figure
\ref{p3:lbs23} because they are unresolved, but OriBN 58 does show
significant depolarization.  Our mapping does not extend to the OriBN
60 core, and further data toward LBS 23 (south) will be needed to
determine whether or not the observed depolarization extends to this
region.

\section{Interpreting Polarization Patterns}
\label{p3:sec:theory}

Polarization vectors in the submillimeter are perpendicular to the
direction of the magnetic field component in the plane of the sky,
appropriately averaged along the line of sight through the cloud.
However, it is generally insufficient to infer the field structure by
simply rotating the polarization vectors by $90^\circ$, since this
procedure implicitly assumes that the direction of the plane of sky
component of the magnetic field is constant.  Magnetic fields in
molecular clouds are almost certainly curved in three dimensions,
which violates this condition for all but the simplest geometries.  A
simple example serves to illustrate that this technique fails.
\citet{fp00c} recently calculated the polarized emission from their
models of filamentary clouds threaded by helical magnetic fields
\citep{fp00a}.  This was done by numerically integrating the
contributions to the Stokes' vectors along lines of sight through the
filament.  They demonstrated that the polarization is always parallel
or perpendicular to the symmetry axis of the filaments in the plane of
the sky, (sometimes with sudden $90^\circ$ flips in orientation at
some radial distance).  Now, if we try to interpret the resulting
polarization maps by simply rotating the polarization vectors derived
from their models, we discover that this approach does not resemble the 
input helical field geometry.  A more sophisticated approach is
generally required to compare candidate models, threaded by
non-trivial 3-dimensional magnetic fields, with the data.  This must
involve direct modeling of polarization patterns.

It is important to note that interpretations of polarization data are
somewhat degenerate.  By varying the magnetic field geometry and the
polarization efficiency of the grains, it is often possible to produce
a qualitatively similar polarization map from more than one field
geometry.  An example of this difficulty is given in $\S$
\ref{p3:sec:NGC2024}, where we present two alternate models of our NGC
2024 polarization map.  It is generally impossible to {\em uniquely}
determine the 3-dimensional structure of the magnetic field from
polarization data alone.  Nevertheless, polarization maps can be used
to provide strong constraints on models of magnetized clouds,
filaments, and cores as follows.  1) On the basis of polarization data
alone, it is often possible to rule out models that are inconsistent
with the data when viewed from all possible orientations.  For
example, in Paper II, we were able to rule out
all models of the \intfil\ in Orion A that are threaded by a purely
poloidal field.  2) All existing magnetohydrostatic models of clouds,
filaments, and cores are axisymmetric.  The assumption of axisymmetry
reduces the set of models to be compared with the data which limits
the allowed configurations.  We find that much of the polarization
degeneracy is lifted when only axisymmetric models, or perturbations
of axisymmetric models (see $\S$ \ref{p3:sec:helical}), are
considered.  We also find that certain field structures produce
identifiable, although sometimes surprising, ``signatures'' in the
polarization data, which can be used as a guide toward developing
detailed models.  3) There are few regions where both polarization
data and Zeeman data exist.  Zeeman data can be combined with the
polarimetry to construct a more nearly unique model than could be
determined by polarimetry alone, since Zeeman observations measure the
line of sight component of the magnetic field.  However, one must be
extremely careful because Zeeman measurements and submillimeter
polarimetry usually trace different components of the gas. An
interpretation that combines Zeeman data and polarimetry implicitly
assumes that the field is related in these components.  We combine
polarization and Zeeman data in $\S$ \ref{p3:sec:expand}, where we
model the NGC 2024 as a shell swept up by an \hii\ region around a
dense ridge.

An alternative approach is to compare the data with polarization maps
predicted from simulations.  The cores arising in simulations are not
generally in equilibium, and the field simulations are not generally
axisymmetric.  This technique has the potential to provide useful
insight into the general statistical behavior of polarization maps for
ensembles of objects.  \citet{pad01} have applied the polarization
modeling technique developed by \citet{fp00d} to simulations of MHD
turbulence.  They find that the relationship between polarization
percentage and submillimeter intensity is a robust statistic, which
can be used to constrain models of grain alignment.  A limitation of
turbulence simulations is that they cannot be easily used to model
specific objects, which is the focus of this paper.

\subsection{NGC 2071}
\label{p3:sec:NGC2071}

NGC 2071 is a massive core that has no associated filamentary
structure.  Nevertheless, there are some strong similarities between
this core and OMC-1, even though the latter is associated with the
massive \intfil\ of Orion A.  There are no Zeeman data toward NGC
2071, so the polarization data alone constrain the field geometry.
The polarization map is similar to the 100 \micron\ polarization map
of OMC-1 made by \citet{sch98}.  In that region, the polarization data
are interpreted as indicating an ``hourglass'' magnetic field, which
developed as the core contracted perpendicular to field lines.  The
contraction is centered on the infra-red source known as KL.  However,
there is an important difference between Figure \ref{p3:n2071} and
Schleuning's 100 and 350 \micron\ maps.  The OMC-1 core exhibits a
flattened structure, whose long axis (in projection) is parallel to
the mean direction of the polarization vectors.  This is precisely
what one would expect for an oblate core threaded by an hourglass
field \citep{mou76,tin88}.  However, the NGC 2071 core is not
flattened at all.  There is no evidence of elongation either parallel
or perpendicular to the outflow in our data (see Figure
\ref{p3:n2071}) or the 50 and 100 \micron\ observations of
\citet{but90}.

The direction of the most powerful outflow of NGC 2071, thought to
originate from the source IRS 3, is shown in Figure \ref{p3:n2071}.
It is aligned at a position angle of 40$^\circ$ (east of north).  The
mean polarization position angle in our data set is $\sim 22^\circ$,
with a dispersion about the mean of $32^\circ$.  In the subset of
vectors toward the peak, the mean position angle is 34$^\circ$ with a
standard deviation about the mean of 33$^\circ$.  The vectors are
generally consistent with with the direction of the outflow.

\citet{hou00} have recently used a new technique, which compares the
line profiles of the coexisting neutral (HCN) and ionic (HCO$^+$)
species, to show that the outflow's magnetic field and its direction
are well-aligned in NGC 2071.  However, our polarization data suggest
a magnetic field that is either orthogonal to the outflow direction,
or strongly toroidal about an axis parallel to the outflow.  In either
case, the magnetic field in the core does not appear to coincide with
the field in the outflow.  The first possibility would be difficult to
reconcile with theory, unless the magnetic field at the small scales
where the outflow originates is unrelated to the large scale field
threading the core.  It would be especially problematic for models of
protostellar formation that favor collapse along field lines,
resulting in outflows aligned with the field of the core.  This
possibility would substantially diminish the role of magnetic fields
in protostellar collapse.  The latter possibility is more intriguing
because it could be reconciled with a class of self-similar outflow
models proposed by \citet{fh96} and elaborated on by \citet{lfh99}.
These models permit a magnetic field which is primarily toroidal in
the dense, partially pressure supported, gas near the equatorial
plane, even far from the central star.  The topology of the field is
quadrupolar in the poloidal plane, so that the field is almost radial
near the symmetry axis defined by the outflow direction and parallel
to the outflow.

\citet{mot01} have compared the SCUBA 850 \micron\ emission from NGC
2071 to pre-existing maps of CO $J=3-2$ and HCO$^+$ $J=4-3$
\citep{che92} and find that up to half the emission in the 850
\micron\ bandpass could be originating in the outflow's line emission.
This could affect the interpretations given above.  \citet{mot01}
estimate that $> 20$\% of the continuum emission at 850 \micron\ could
arise from the CO $J=3-2$ line.  According to theory, in the case of
high density gas, the linear polarization produced in spectral lines
should be parallel to the magnetic field.  Even if the magnetic fields
giving rise to the dust and line polarizations were perfectly aligned,
they would contribute orthogonal vectors to the net polarization.
Thus, the usual interpretation of submillimeter continuum polarization
as being orthogonal to the plane-of-sky component of the magnetic
field could be incorrect in regions where the CO emission is strong
enough to dominate the continuum flux from dust.  If our polarization
data were dominated by line emission, then the inferred field
direction would align to within 10$^\circ$ of the CO outflow.
However, where the CO emission is optically thick, the maximum
polarization contributed by the outflow is on the order of 1\%, as is
the case for CO $J=3-2$ emission from the outflow of FIR 5 in NGC 2024
\citep{gre01a}.  Therefore, when dust polarizations exceed 1\%, they
are unlikely to be dominated by CO line contamination (J.\ Greaves,
private communication).  Therefore, we conclude that the strong
polarizations measured toward this core are dominated by dust emission
and hence do not infer a magnetic field aligned with the outflow
direction.

Finally, we note that the orientation of the outflow relative to the
core's magnetic field may turn out to be inconsequential.  NGC 2071 is
a massive core forming many stars.  This is very different from the
case of low mass cores forming isolated stars, to which most of the
existing theory applies.  Within several star-forming regions,
multiple outflows of various orientations are observed, i.e., OMC-2/3
in Orion A \citep{aso00,yu00}.  If the collapse of massive cores is
really dominated by the mean magnetic field, then one might expect a
rough alignment between all outflows originating from the same core.
Since this is not observed in NGC 2071 \citep{eis00}, it is unclear
how relevant the large-scale field is in determining the orientation
of outflows.
 
\subsubsection{An Estimate of Magnetic Field Strength}

\citet{cf53a} demonstrated that when a mean field direction can be
defined, the magnetic field strength can be derived using the
dispersion in polarization angles, $\sigma_\theta$, the dispersion in
velocity along the line of sight, $\sigma_{v_{los}}$, and the mean
density, $\rho$, in the region local to the polarization and velocity
data.  In cases of complex, ordered field geometries, it is no
easy task to assign a mean field direction based on polarization data
(see \S \S \ref{p3:sec:LBS23N} and \ref{p3:sec:NGC2024} below).
However, in NGC 2071, we can apply the method of \citet{cf53a}
assuming that the field threading the core is relatively straight.
The mean field strength, $\meanBfield$, is given by
\begin{equation}
\langle B\rangle = \left [ 4 \pi \rho \frac{(\sigma_{v_{los}})^2}{(\sigma_\theta)^2} \right ]^{1/2}.
\label{meanB}
\end{equation}

\noindent The NGC 2071 core was mapped in NH$_3$ (1,1) emission at
high velocity resolution by \citet{tak86}.  The ammonia emission
extends over 2\arcmin\ from the dust peak location and shows more
elongation than the dust map.  \citep{tak86} identified three
components in the line profiles from across the core.  The narrowest
feature is described as a ``spike'' in the profile and is seen 
over the entire mapped area.  This emission is attributed to the core itself
(as opposed to any outflowing gas).  The FWHM of this component
is estimated to be 1 \kms, which corresponds to a dispersion of
approximately 0.5 \kms.  Ammonia traces relatively high density gas
($n(H_2) \ge 10^4$ cm$^{-3}$), so we use $10^4$ cm$^{-3}$ as an
estimate of the mean density across the core.  As discussed above, the
dispersion in polarization angles at 850 \micron\ is 33$^\circ$, which
corresponds to 0.58 radians.  Substitution of these values into
equation (\ref{meanB}) yields $\meanBfield \approx 56 \mu$G which is
consistent with what would be expected within a dense core.

\citet{hei01} propose an amendment to the Chandrasekhar and Fermi
model which yields an estimate of the rms field strength.  In the case
of equipartition between the magnetic and kinetic energies, the total
magnetic energy is given by
\begin{equation}
\langle B^2 \rangle = 4 \pi \rho \frac{(\sigma_{v_los})^2}{(\sigma_\theta)^2} [1+(3\sigma_\theta^2)]
\label{Brms}
\end{equation}

\noindent which gives similar results to equation (\ref{meanB}) when
the field is highly ordered.  For NGC 2071, we find that the ratio of
the total magnetic energy, $\langle B^2 \rangle$, to the mean field
energy, $\langle B \rangle^2$, is 2.  Using the values for NGC 2071
above, the rms field is estimated to be 78 $\mu$G, a factor of 1.4
times greater than the mean field.

\citet{cru99a} compared the densities within star-forming clouds with
the line-of-sight field strengths measured using the Zeeman effect on
molecular lines.  Based on his compilation of all existing data, he
concluded there was no substantial evidence against equipartition in
star-forming clouds.  The line-of-sight field strength, $B_{los}$, can
be extrapolated from the mean field strength based on a statistical
average over all potential inclinations of the field to the line of
sight.  We estimate $B_{los}$ to be roughly the mean field divided by
2, or 28 $\mu$G.  Comparing this field strength to the density within
the NGC 2071 core of $10^4$ cm$^{-3}$ using Table\ 1 of
\citet{cru99a}, it is clear that this value is comparable to
measurements of the Barnard 1 dark cloud core, but substantially less
than the more massive M17 cloud. 

Based on this comparison, we conclude that, while the
Chandrasekhar-Fermi method does give a value of mean field strength
consistent with those measured directly using Zeeman splitting, it
also predicts that approximately half the magnetic energy does not
reside in the mean field component.  These results are not surprising
given the fact that an ordered pattern is observed in the polarization
data which is not straight.  Thus, we should not be too quick to
attribute the rest of the magnetic energy to turbulence.  However,
direct measurement of Zeeman splitting of OH lines at several
positions in NGC 2071 would provide a further test on the validity of
a simple, mean field geometry within this cloud.

\subsection{LBS 23N}
\label{p3:sec:LBS23N}

The polarization structure of the LBS 23N region is somewhat less
orderly than that of either NGC 2071 or NGC 2024, and there are no
Zeeman data toward this region. However, there remains sufficient
structure in the polarization map to warrant some comparison to
filamentary models.  The LBS 23N region is a linear string of roughly
equally spaced condensations, which might be the result of
fragmentation due to gravitational instabilities (see \citet{cf53b},
\citet{nak93} and \citet{fp00b} for a discussion of gravitational
instabilities in magnetized filaments).  The overall orientation of
the polarization vectors is along the filament, with some degree of
scatter in position angle.  The mean position angle in the 70 vectors
is $8.7^\circ$ with a standard deviation of $5.0^\circ$.

We offer three possible explanations for such a pattern.  The first
possibility is that the original filament is threaded by a magnetic
field transverse to the symmetry axis.  This magnetic field geometry
would reproduce the orientation of the majority of the
polarization vectors.  However, it is not easy to understand how such
a field geometry could be associated with a filament or linear string
of cores collapsed in three dimensions, since a transverse field would
more likely be associated with a sheet-like distribution of material.
This geometry also cannot readily explain the 90$^\circ$ flip seen in the
polarization vectors toward the southern end of the filament.

Alternatively, \citet{fp00c} proposed that such a pattern could be
explained as the result of a helical field dominated by the toroidal
field component (Type 1, according to their classification) as
observed in the OMC-3 filament of Orion A.  Helical field geometry can
create polarization patterns where vectors flip by $90^\circ$ at the
interface between poloidally and toroidally dominated regions, which
could occur if the field threading the southern part of the filament
is less tightly wound than in the north.  Such a flip is suggested by
our data at the southern end of the polarimetry map.  We note
that these data provide only sparse coverage of this region and are
the lowest signal-to-noise data of all three regions.  Modeling of
this region will require observations of the southern part of the
filament.

Finally, Paper II presents a model for two crossed filaments,
unconnected but overlapping in projection on the plane of the sky.  In
such a case, the vectors in the overlap region can be dominated by one
of the two filaments.  The east-west filamentary structure is
identified in Figure \ref{p3:lbs23}.  As for OMC-3, this scenario can
be tested by more extensive mapping into LBS 23 south.

\subsection{NGC 2024}
\label{p3:sec:NGC2024}

The overall appearance of NGC 2024's submillimeter emission is that of
a short filament, which is broken into three main condensations.  The
polarization vectors show a complicated but extremely well-ordered
pattern throughout the region.  An interesting feature of the map is a
rather abrupt change in orientation across the central ridge of the
filament.  This feature cannot be reconciled easily with a purely
poloidal field along the filament axis, since that geometry would
produce polarization vectors perpendicular to the axis, irrespective
of the inclination of the filament.  \citet{fp00c} found that
polarization vectors can flip by $90^\circ$ across the central axis of
a filament threaded by a helical field, when the filament is inclined
at an angle relative to the plane of the sky.  However, the
polarization vectors in their model are always aligned either parallel
to or perpendicular to the filament.  This is not the case in
our map, where the vectors form a ``V'' like pattern, with the apex
located on the central ridge of the filament.  However, we show below
that a simple extension to their model can account for the
qualitative features of this map.

The interpretation of the polarization pattern is constrained by
measurements of the line-of-sight magnetic field component, \blos\
(sometimes denoted $B_{||}$).  \citet{crtg99}, using Zeeman splitting of OH
absorption lines, detected a peak in \blos\ of almost 100 $\mu$G at a
position west of FIR 4.  The field strength declines smoothly from
this southwestern position in their map to the northeast, where it
drops to zero (cf.\ their figure 3).  OH samples gas densities of
approximately $10^4$ cm$^{-3}$, which are substantially lower than the
volume densities estimated along the NGC 2024 molecular ridge.  The
extended filamentary structure in which the cores are embedded have
$<n_H> \approx 5 \times 10^6$ cm$^{-3}$, while the cores themselves
have $<n_H> \approx 2 \times 10^8$ cm$^{-3}$ \citep{mez92}.  Thus, it
is not certain that the field being traced by the Zeeman data threads
the same dense gas whose magnetic field is traced by our polarization
measurements.  However, the fact that the positions of low
polarization percentage in both 100 \micron\ and 850 \micron\
correspond closely with the maximum in line-of-sight field could
support a correlation in the fields traced by the two techniques,
since low polarizations could indicate a region where \blos\ is
dominant.  Conversely, the low polarization could be a produced by
variations in alignment or grain properties and thus be unrelated to
\blos.  Keeping this uncertainty in mind, we offer two theoretical
models for this pattern.

\citet{sch91} measured NH$_3$ and CS gas toward NGC 2024 and found
distributions of dense gas that closely mirrored the pattern of dust
condensations.  This supported later conclusions that the dense
ridge is located on the far-side of the \hii\ region
\citep{cc96,crtg99}, as originally schematically illustrated by
\citet{bar89}.  We illustrate the most likely configuration in Figure
\ref{p3:mycartoon1}.  In order for OH to be observed in absorption,
there must be a source of background continuum emission.  Thus, either
the OH Zeeman data trace the field in the foreground dust lanes (with
the observable \hii\ region as the continuum source) or the OH
absorption is occuring in the denser molecular cloud, thus requiring a
second source of continuum emission behind the ridge.  For instance,
the \hii\ region could have penetrated to different depths of the
cloud due to density variations.  The latter scenario is favored by
\citet{crtg99} with the ridge located on the far side of the \hii\
region.

In the first case, the Zeeman splitting OH absorption data may be
sampling an envelope of gas either in the foreground of the \hii\
region itself, or simply at the boundary between the \hii\ region and
the very dense molecular cloud behind it, regardless of the location
of the dense ridge.  In this case, the Zeeman data are unrelated to
the polarization data and would not provide a constraint.  In the
second case, we assume that although the OH data sample a less dense
region of gas, the measurements are sampling the same field geometry
in both regions.

\subsubsection{A Helical Field Model}
\label{p3:sec:helical}

Perfectly straight, infinite filaments are an idealization, which may
need to be relaxed somewhat when comparing models with real data.
Figure \ref{p3:model} shows a finite segment of a filament model from
\citet{fp00a}, which we have ``bent'' into a circular arc using a
simple mathematical transformation that maps planes parallel to the
axis of the filament into concentric cylinders of some specified mean
radius of curvature, orthogonal to the filament axis.  The density and
magnetic field are transformed self-consistently, using the Lagrangian
formulation of the MHD induction equation and the equation of mass
conservation (see Parker for example):
\begin{eqnarray}
\frac{\bf B}{\rho} &=& {\bf J} \frac{\bf B_0}{\rho_0} \\
\rho &=& \frac{\rho_0}{| {\bf J} |},
\label{eq:lagrange}
\end{eqnarray} 

\noindent where ${\bf J}$ is the Jacobian of the coordinate
transformation and $|{\bf J}|$ is its determinant.  We will thoroughly
discuss the details of our modeling technique in a future paper.

The polarization structure of the theoretical map reproduces most of
the qualitative features of Figure \ref{p3:n2024}.  On the other hand,
models that are threaded by purely poloidal fields do not agree with
the polarization data.  Note that the model shown is a preliminary
result and should not be regarded as our final, best fit to the data.
Nevertheless, we have provided the exact parameters of this model in
the caption of Figure \ref{p3:model}.

It is interesting that a bend in the filament is required to reproduce
the polarization pattern shown in Figure \ref{p3:n2024}.  If the
filament is bent so that the source FIR 5 is closest to the expanding
ionization front of the \hii\ region, then this could explain why it
is the most evolved source along the ridge.  The outflow from this
source is the most powerful in the region \citep{ric92}, while
evidence for less evolved outflows can be found in the sources FIR 4
and FIR 6 \citep{cc96}.

\subsubsection{Expansion of the \hii\ Region around A Dense Filament}
\label{p3:sec:expand}

The expected Zeeman pattern from a helical field is a reversal in
orientation of \blos\ across the symmetry axis of a filament.  If the
field threading the filament is related to the field in the lower
density gas, then the helical field interpretation would not be
consistent with the observations of \citet{crtg99} toward the northern
portion of the NGC 2024 dense ridge.  In order to reconcile the Zeeman
and polarization data, we suggest that the magnetic field has been
distorted from its original structure by the expansion of the \hii\
region, as shown in the schematic diagram of Figure \ref{p3:mycartoon2}.  The
field may be pushed out on the surface of the ionization front, but
will bend around the dense, filamentary system of cores comprising the
NGC 2024 region.

The asymmetric Zeeman data requires one of the following
geometries: either the ionizing source is in front of and to the
west of the dense ridge; or the density of the molecular cloud itself
must increase to the east, thereby inhibiting the expansion of the
\hii\ region on that side \citep{bar89}.  The contours of Figure
\ref{p3:2pols} show that the dust extends further to the east of the
ridge than to the west.  If the source of the ionization front is to
the west of the molecular ridge, then the vectors seen from our line
of sight will stretch across the front of the ridge and to the east
nearly in the plane of the sky (i.e., the \blos\ component is zero).
To the west of the molecular ridge, such a source would produce an
ionization front which would be bending around the ridge, with
components \blos\ and \bperp\ both non-zero.

However, there is substantial evidence that the dense cores and
high density gas lie behind the \hii\ region, as shown in Figure
\ref{p3:mycartoon1}.  Although \citet{bar89} calculated the ionizing
radiation detected in the nebula to require the IRS 3 source to be an
O9 (or earlier) star, they conclude this to be unlikely unless the
extinction toward IRS 3 is 4.1 mag in $K$, a factor of 2 greater than
the average extinction derived toward that star.  Instead, they
suggest that an ensemble of B stars must have formed the \hii\ region.
If each star has formed a bubble of ionized radiation around it, as
IRS 2 has produced the Eastern Loop (see Figure \ref{p3:mycartoon1}),
then the cumulative effect could produce the \hii\ region.

In order for this picture to be physically reasonable, the ionization
front must be interacting with the molecular ridge.  \citet{gau92}
detected 1.3 cm emission at 3\arcsec\ resolution located at south of
FIR 4 (their NH$_3$ 4) and north of FIR 5 (their NH$_3$ 5).  North of
the FIR 5 source, the 1.3 cm emission arises from a dense, ionized
interface between the molecular cloud (and ridge of cores) and the
\hii\ region.  Ionized emission near FIR 4 has a compact component
associated with the dense ridge.  Additionally, higher NH$_3$
temperatures, larger line widths and smaller abundances suggest that
the FIR 4 and FIR 5 sources are nearest the ionization front of the
\hii\ region.  \citet{sch91} show that ionized gas in fact fills the
bay in the molecular bar surrounding IRS 3 (see their figure 7), and
that the ridge is effectively separated into two subregions by the
ionization front of the \hii\ region.  We proceed on the basis that
the ridge existed prior to the formation of the \hii\ region.

The polarization model shown in Figure \ref{p3:bentfield} is based on the
idea that the magnetic field has been swept up in a shell during the
expansion of the \hii\ region and stretched over the surface of the
dense filamentary system of cores embedded in the wall of the
molecular cloud.  The stretching of the field in this scenario would
tend to make the field quite uniform by the time the shell reached the
system of cores.  The structure of the magnetic field is idealized as
follows.  We assume that the field is swept into a sheet with a
Gaussian planar distribution of the form 
\begin{eqnarray}
\rho &=& \rho_c e^{-\frac{(y_0-y_c)^2}{\delta y^2}} \\ 
B_0 &=& B_{x0} e^{-\frac{(y_0-y_c)^2}{\delta y^2}} \hat{x} + B_{z0} e^{-\frac{(y_0-y_c)^2}{\delta 
y^2}} \hat{z}
\label{p3:eq:fieldStruct}
\end{eqnarray}
when it first encounters the embedded filament, where $\hat{y}$ is
along the line of sight, $y_c$ is the offset of the field's center
from the filament axis, $\delta y$ is the width of the field region
and $B_{x,0}$ and $B_{z,0}$ are respectively the maximum, central
values of the $\hat{x}$ and $\hat{z}$ field components.  The magnetic
field threading the shell is generally not aligned with respect to the
filament axis, which is oriented in the $\hat{z}$ direction.  The
filament is assumed to be an unmagnetized, \citet{ost64} filament,
which we have truncated at a radius of $10^C~r_0$, where $r_0$ is the
radial scale of the filament given in \citet{fp00a}.  This radial
scale basically defines the radius where the inner, nearly flat
density profile begins to steepen toward the $r^{-4}$ profile that the
density obeys at $r \gg r_0$.

We wrap the field threading the shell around the filament by applying
a coordinate transformation of the form \be x=x_0, ~~y=y_0+r_0 (x/w)^2, ~~z=z_0
\label{p3:eq:transform}
\ee to the planar field giving by equation (\ref{p3:eq:fieldStruct}).
The front of the shell is defined by the surface $y_0=-10^C r_0$, so
that it just touches the filament.  This transforms the originally
planar sheet into a parabolic shell, wrapped around the filament,
which is centered on the origin.  The magnetic field and density are
self-consistently distorted using the Lagrangian forms of the MHD
induction equation and the equation of mass conservation, as given by
equation \ref{eq:lagrange}.  The determinant $|{\bf J}|$ of the
transformation is unity, so that the density is unchanged.

The Zeeman data of \citet{crtg99} suggest that the magnetic field is
approximately parallel to the surface of the molecular cloud east of
the filament, and strongly bent into the line of sight on the west.
This is also supported by \citet{bar89}, who suggest that the
molecular cloud may be denser east of the ridge than to the west,
since the field would be swept further back on the side of lower
density.  Thus we rotate our field configuration by $20^\circ$
counter-clockwise (looking toward the origin from the ${\hat z}$ axis)
to mimic this asymmetry.  The inclination angle that best reproduces
the observed polarization map is $60^\circ$ into the plane of the sky.
We apply a final rotation of $-15^\circ$ in the plane of the sky to
better agree with the observed orientation of the filament.  The
resultant polarization pattern is shown in Figure \ref{p3:bentfield}.
This figure also shows a map of the density-weighted average of the
line of sight component of the magnetic field (arbitrary units) for
comparison with the Zeeman data.

This model is in better agreement with the Zeeman data for NGC 2024
than the helical field model given in $\S$ \ref{p3:sec:helical}.
However, our magnetized shell model has several shortcomings.  The
best-fitting polarization pattern that can be obtained from this type
of model, with the Zeeman data regarded as a constraint, is of
noticeably poorer agreement than the helical field model.  The shell
model reproduces the general features of the map, but the helical
field model does a better job accounting for the detailed structure.
The shell model also predicts a dense ridge of emission on the west
side of the filament, where the parabolic shell plunges into the plane
of the sky.  It is noteworthy that the filament in this model is
unmagnetized, and therefore does not contribute to the polarized flux.
Hence, it is necessary to put a considerable amount of mass into the
shell, so that the polarization percentage, defined as the polarized
flux divided by the total intensity, is sufficiently high across the
filament to account for the observations.  The central density of the
shell is 11.4\% of the central density of the filament, and the
shell's thickness is $10^C/3$.  A side-effect of this dense shell is
that the emission away from the filament is non-negligible.  Both the
polarization and the intensity in Figure \ref{p3:bentfield} are
clipped at 5\% of their respective maximum values.  Nevertheless,
polarized emission above these levels is found in the background, in
contrast to the helical field model, where this clipping procedure
eliminates the flux away from the filament.

The depolarization observed in NGC 2024 is easily understood in this
model.  The filament itself does not contribute to the polarized flux,
since it is unmagnetized.  Instead, a uniformly polarized foreground
sheet produces the polarization.  Where the total intensity is high
(i.e., along the filament), the polarization percentage declines, as
we observe.

A limitation of this model is that the filament is unmagnetized,
and therefore obeys the $r^{-4}$ density profile predicted by Ostriker
(1964).  This density gradient is known to be too steep to account for
the observations in other filaments (\citet{alves98}; \citet{lal99};
\citet{joh99}).  A magnetic field threading the filament could be
included in this model, in principle, but this would make it
substantially more complex.

A third type of model may also be worth considering.  We have assumed
in the above model that a shell swept up by the \hii\ region wrapped
around a dense filament, pre-existing in the cloud behind the \hii\
region.  One could imagine an alternate scenario involving the
filament fragmenting out of the shell.  The fastest growing unstable
modes of fragmentation of a magnetized sheet result in filements
oriented either parallel or perpendicular to the magnetic field,
depending on the thickness of the shell \citep{nagai}.  In either
case, the geometry of the field probably could not account for the
complex structure seen in our polarization map.  We also note that a
fragmenting shell would most likely produce more than one filament,
and no others are observed in the vicinity of NGC 2024.

\subsection{Further Observational Tests of the Helical Model}

Although quantitative comparisons of the filamentary
helically-threaded models to 850 \micron\ data (and where applicable,
Zeeman data) will constrain which model can reliably
reproduce our observations, different observation tests will
eventually be possible.  For example, observations a) on larger
spatial scales, b) over a range of wavelengths, or c) at higher
resolutions will each test the helical field models.

If filaments are threaded by helical fields, then the poloidal
currents associated with those fields must connect to the larger scale
magnetic field of the molecular cloud itself. The polarization
signatures on these larger scales will be detectable once more
sensitive submillimeter cameras (e.g. SCUBA-2) are built. Higher
senstitivity is only one goal of the next generation of detectors; the
most important advance will come with the elimination of the need to
chop on the sky.  This will enable observations of polarized emission
from faint, relatively featureless dust structures on scales larger
than the detector field of view.

The second extended test of helical field geometries can be made with
high resolution, interferometric, mapping.  The helical field geometry
predicts that depolarization should persist down to very small scales.
In practice, such maps are hard to interpret since current
interferometers are limited to observations of cores where star
formation is already occurring.  Ideally, one would like to observe
filaments at equally high resolution, but this goal is hindered by
sensitivity limits and the filtering of large scale structure beyond
some limit imposed by the spacing of the array.  While sensitivity
will continue to improve, the filtering effect is not so easily
overcome where short spacings information is not present.

Perhaps the most immediately tangible way to further test helical
field models is the observation of regions over a range of
wavelengths.  If one were to observe shorter wavelengths until sources
become optically thick, then the measurements could be used as a depth
probe of the magnetic field.  Although grain effects can produce
changes in polarization percentage \citep{hil99}, changes in position
angle are attributed to the magnetic field.  If a field is indeed
helical, the vectors should change orientation relative to a straight
filament, from alignment with the filament axis to orthogonality as
wavelength decreases.  The future SOFIA polarimeter, operating below
200 \micron\ will be particularly useful for testing this aspect of
the models.

\section{Summary}
\label{p3:summary}

The polarization patterns observed in Orion B are dominated by orderly
structure on scales at least as large as the areas mapped (with the
possible exception of the LBS 23N region).  This result can only be
consistent with magnetic fields that are ordered on similar scales or
larger.  The filamentary clouds, NGC 2024 and LBS 23N, exhibit
polarization patterns that are inconsistent with purely poloidal
magnetic fields threading the filaments, or with fields simply
threaded transversely to them.  Curved field lines are necessary to
model the polarization data in NGC 2024.  The fact that the
polarization pattern is symmetric about the dense ridge of cores
suggests that there is a correlation between the presence of dense gas
and the exhibited polarization pattern.

The polarization systematically decreases with intensity for all three
regions mapped.  This result is in agreement with Paper II, which
notes depolarization toward the axis of the \intfil\ in
OMC-3, as well as most submillimeter and millimeter observations of
star-forming regions (see \citet{wga00} for a review).  This
``polarization hole'' effect is a {\em global} property of our maps,
and not just a local phenomenon that occurs for a few pixels near the
very brightest emission.

NGC 2071 is a massive core forming multiple protostars.  Its
polarization pattern is ordered and qualitatively similar to that of
OMC-1, which is an even more massive core in Orion A.  In OMC-1,
\citet{sch98} interpreted their polarization data as being due to an
hourglass magnetic field, with the field lines pinched due to the
collapse of the core.  This interpretation is consistent with the
flattening observed along the inferred field lines in OMC-1.  On the
other hand, NGC 2071 does not show any flattening, which is
inconsistent with an interpretation of the magnetic field threading
this core as a dynamically significant hourglass field.  We note that
if the vectors of NGC 2071 were rotated to infer a net field
direction, this direction would be {\it perpendicular} to the most
powerful outflow in the region.  Alternatively, our map could be
interpreted as resulting from a field that is predominantly toroidal
about the axis of symmetry.  Although CO $J=3-2$ emission can
contaminate SCUBA 850 \micron\ data, we estimate it can contribute a
maximum of 1\% of polarized emission when optically thick.  Since much
higher polarizations are measured, we conclude that the polarizations
measured in NGC 2071 are dominated by dust emission.  Application of
the Chandrasekhar \& Fermi method to this core yields a mean field
strength estimate of 56 $\mu$G, and we find that half the magnetic
energy is accounted for by the mean field component.

LBS 23N exhibits the least orderly polarization patterns of the three
regions studied.  Most of the polarization vectors are aligned in a
north-south orientation, particularly to the east of the cores.  The
cores themselves are significantly depolarized.  Near the southern
boundary of the map, the vectors rotate by $90^\circ$ to an east-west
orientation.  This abrupt change might be explained either by an
extension of the Fiege \& Pudritz model (see \S \ref{p3:sec:LBS23N}),
or by a smaller filament orthogonal to the main filament.  More data
to the south of LBS 23N would be needed to distinguish between these
possibilities.

The NGC 2024 ridge of cores presents the most interesting polarization
data of this paper.  The polarization is strong everywhere except the
cores, which are significantly depolarized.  Our map agrees very well
with the 100 \micron\ polarization data obtained with the KAO Stokes
polarimeter \citep{dot00}, even though these wavelengths likely probe
different dust temperatures.  Interestingly, the region between FIR 4
and FIR 5 exhibits diminished intensity but no polarization is
detected there.  We are able to successfully model the polarization
pattern using a helical field geometry threading a curved filament.
However, given the overall geometry of the region -- an \hii\ region
expanding into the molecular cloud in which the cores are embedded --
we slightly prefer a model in which the ionization front has swept up
the magnetic field and is now stretching it around the ridge of dense
cores.  Since this model contains an unpolarized filament, it may be
oversimplified and a more complete, complex picture will include both
the polarized ionization front and a polarized filament.

We note that our analysis of the NGC 2024 polarization data have been
aided immensely by the existence of Zeeman maps of the line-of-sight
field strength toward part of this region published by \citet{crtg99}.
Unfortunately, very few regions have been successfully mapped using
Zeeman splitting \citep{cru99a}.  This is an area where future
technological advances in instrumentation would greatly improve our
understanding of magnetic fields in molecular clouds.

\acknowledgements

The authors would like to thank J.~Greaves and T.~Jenness for
assistance during observing and with data reduction.  We also extend
thanks to our fellow members of this JCMT key project, particularly
C.~Wilson and R.~Pudritz for feedback on this manuscript, and our
referee for a thorough and insightful review.  The research of BCM has
been supported through grants from the Natural Sciences and
Engineering Research Council of Canada.  BCM acknowledges funding from
Ontario Graduate Scholarships.  JDF acknowledges support from a
postdoctoral fellowship, jointly funded by the Natural Sciences and
Engineering Council and the Canadian Institute for Theoretical
Astrophysics.

\appendix
\section{Polarization Data}
\label{p3:appendixA}

\begin{figure}
\vspace*{8cm}
\includegraphics{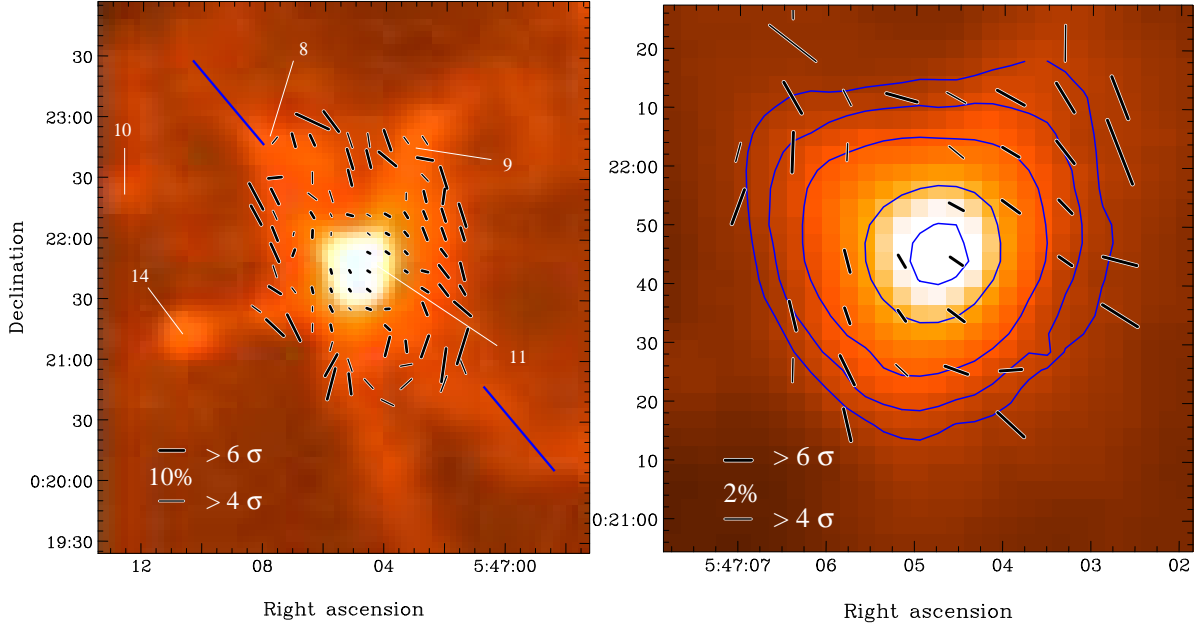}
\caption{850 \micron\ polarization of NGC 2071.  NGC
2071IR greyscale from \citep{mit01} overlain with polarization
vectors.  The vectors are binned to 3 pixels (9\arcsec) so each vector
is separated by just over half the 850 \micron\ beamwidth (14\arcsec~
at the JCMT).  Vectors are only plotted where uncalibrated $I> 2$\% of
the source peak, polarization percentage exceeds 1\% and the absolute
uncertainty in polarization percentage is no greater than 1.5\%.  All
vectors plotted have signal-to-noise in polarization percentage,
$\sigma_p >4$ (and therefore an uncertainty in position angle,
$d\theta < 7.2^\circ$), but bold vectors have $\sigma_p >6$ ($d\theta
< 4.8^\circ$).  Over the entire region, the mean position angle is
$22^\circ$ while the standard deviation is 32$^\circ$. The mean
polarization percentage is 5.1\% with a standard deviation of 3.3\%.
Cores are labelled as in \citet{mit01}, and blue lines mark the mean
direction of the outflow from IRS 3.  {\it Right:} Details of the
central core (\# 11 of \citet{mit01}) with the same vector cutoffs
except $I>12$\% of the peak flux.  Blue contours trace the surface
density in unpolarized 850 \micron\ intensity.  All coordinates are
J2000.}
\label{p3:n2071}
\end{figure}

\begin{figure}
\vspace*{16cm}
\includegraphics{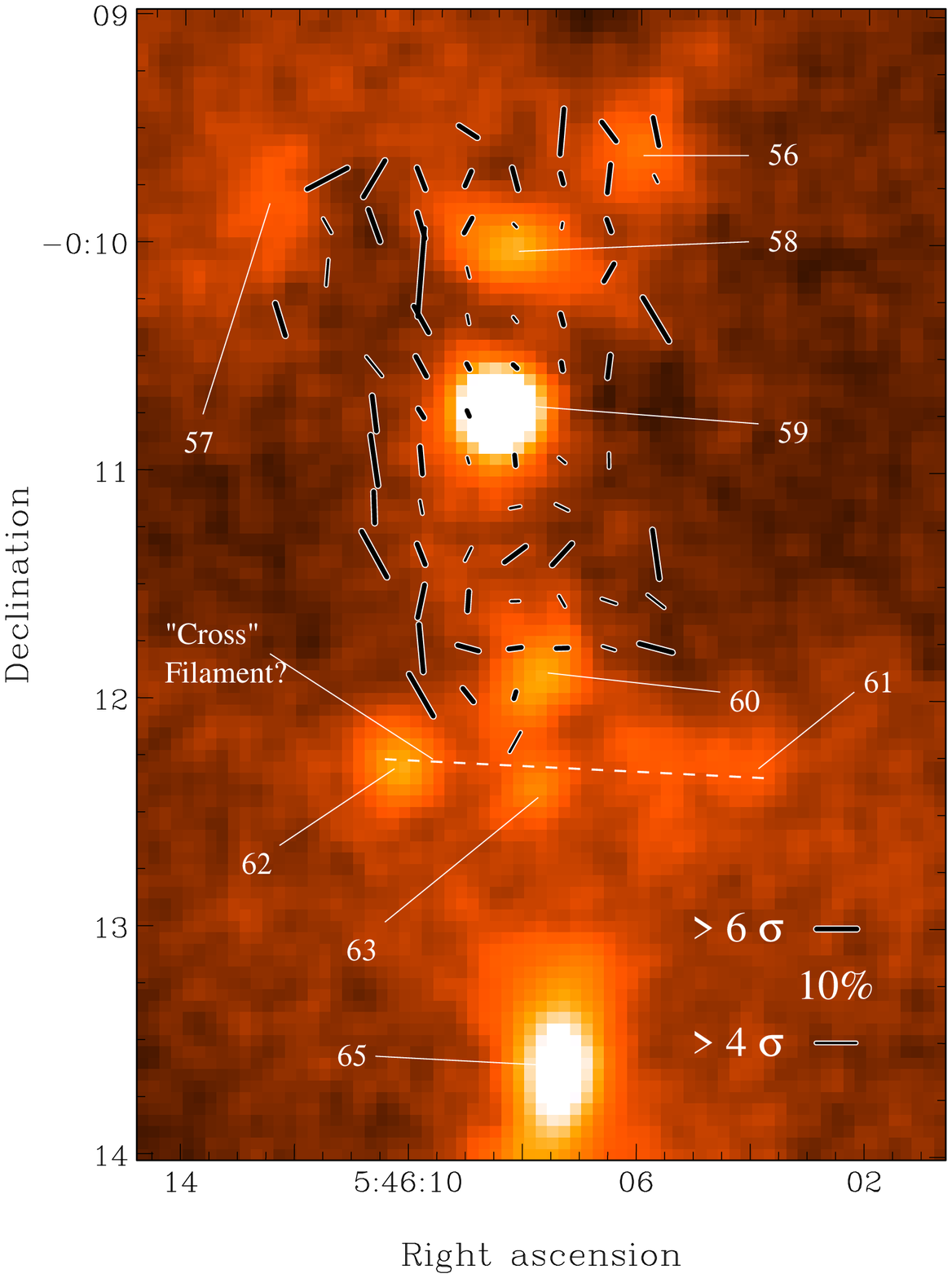}
\caption{850 \micron\ polarization of LBS 23N.  LBS 23
greyscale is taken from \citet{mit01}.  The polarization vectors
overlain have been binned to 12\arcsec.  Vectors are plotted where $I>
4$\% of the peak of core 59, $p>1$\%, $dp<1.5$\% and $p/dp>4$
($d\theta < 7.2^\circ$).  Bold vectors show those where $p/dp>6$
($d\theta < 4.8^\circ$).  The core identifications are taken from
\citet{mit01}.  The mean polarization percentage measured is 5.6\%
with a standard deviation of 3.8\%.  A potential ``cross'' filament
could be oriented east-west and has been indicated by the presence of
a dashed line.  Coordinates are J2000.
}
\label{p3:lbs23}
\end{figure}

\begin{figure}
\vspace*{16cm}
\includegraphics{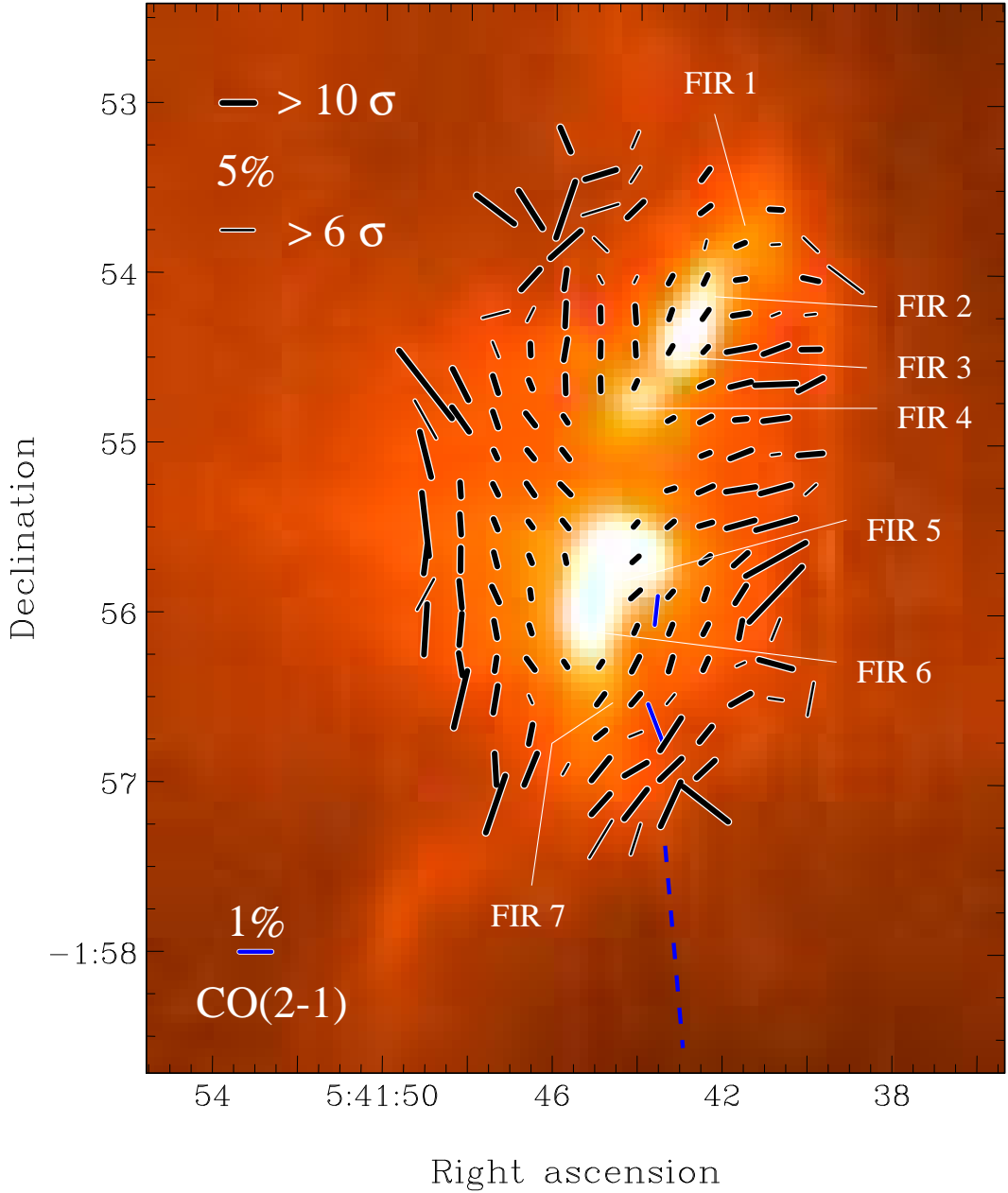}
\caption{NGC 2024 850 \micron\ polarization.  NGC 2024
greyscale overlain with polarization 12\arcsec\ binned polarization
vectors, almost the JCMT beamwidth of 14\arcsec\ at 850 \micron.
Vectors are plotted where $I> 2.5$\% of the FIR 5 peak, $p>1$\%,
$dp<1$\% and $p/dp>6$ ($d\theta < 4.8^\circ$).  Bold vectors show
those where $p/dp>10$ ($d\theta < 2.9^\circ$).  Overall, these data
represent the highest signal to noise polarization detections in our
data set.  The mean percentage polarization is 3.4\% with a standard
deviation of 2.3\% in 159 vectors.  The submillimeter cores are
labelled as in \citet{mez88}.  We have indicated the orientation of
the FIR 5 unipolar outflow by a blue dashed line, and the CO $J=3-2$
line polarization vectors have been plotted (in blue) as measured by
\citet{gre01a}.  Coordinates are J2000.}
\label{p3:n2024}
\end{figure}

\begin{figure}
\vspace*{12cm}
\includegraphics{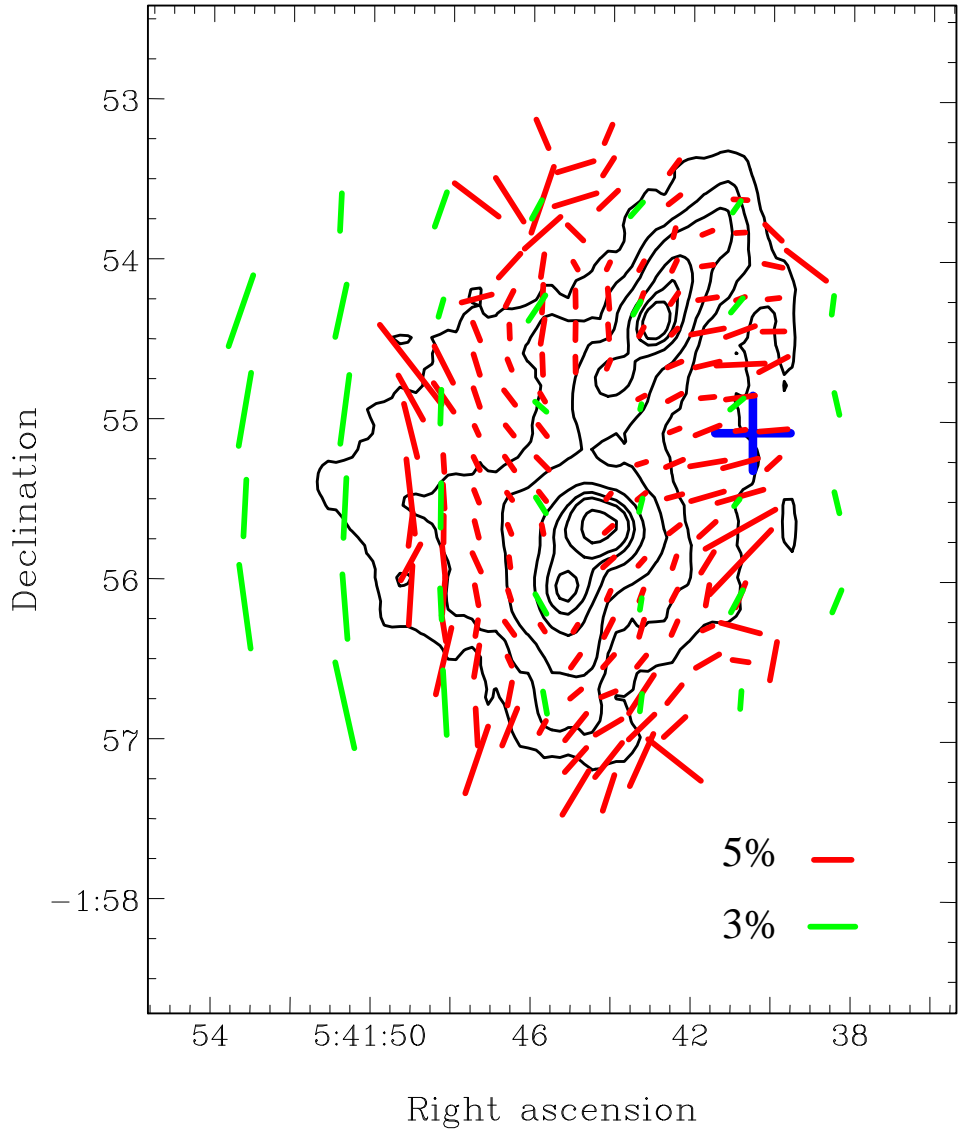}
\caption{Comparison of Polarizations at 100 \micron\ and
850 \micron. Comparison of the polarization patterns at 850 \micron\
(red) and 100 \micron\ (green) reveals a strong consistency in
position angle.  The polarization percentages are considerably lower
for 100 \micron, possibly indicating that fewer dust grains are
sampled within hotter dust or, more probably, that the dust grain
alignment is lessened at higher temperatures.  The 850 \micron\
contours are plotted at 80, 90, 95, 98, 99, 99.5 and 99.9 percentiles.
A blue cross marks the location of the maximum line-of-sight magnetic
field as measured by \citet{crtg99}.}
\label{p3:2pols}
\end{figure}

\begin{figure}
\vspace*{15cm}
\includegraphics{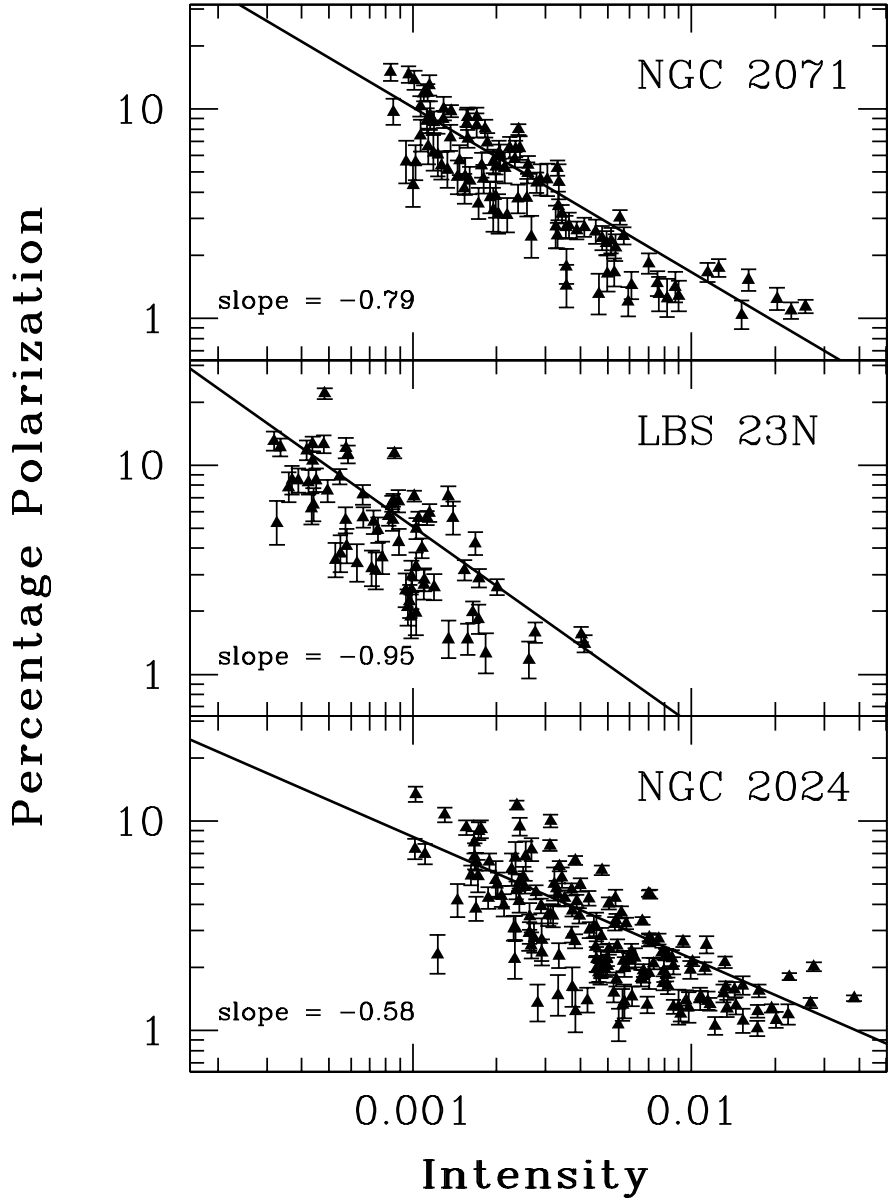}
\caption{The Depolarization Effect in Orion B. The
depolarization effect typically seen in protostars and extended
regions (such as OMC-3 in Orion A) is also observed in NGC 2071, NGC
2024 and LBS 23N.  For all three regions, the brightest positions have
the smallest polarizations, exhibiting $< 2$\% whereas the faintest
regions show polarizations approaching and exceeding 10\%.  Decreased
polarization percentage is observed for a large range of $I$ values in
NGC 2024.  This region's cores also contain the largest areas where
depolarization is total, or the levels of polarization are so low
(i.e., $<1$\%) as to be consistent with zero.  Extremely high volume
densities (on the order of $10^8$ cm$^{-3}$) have been estimated in
NGC 2024 \citep{mez88}.}
\label{p3:depol}
\end{figure}

\begin{figure}
\vspace*{15cm}
\includegraphics{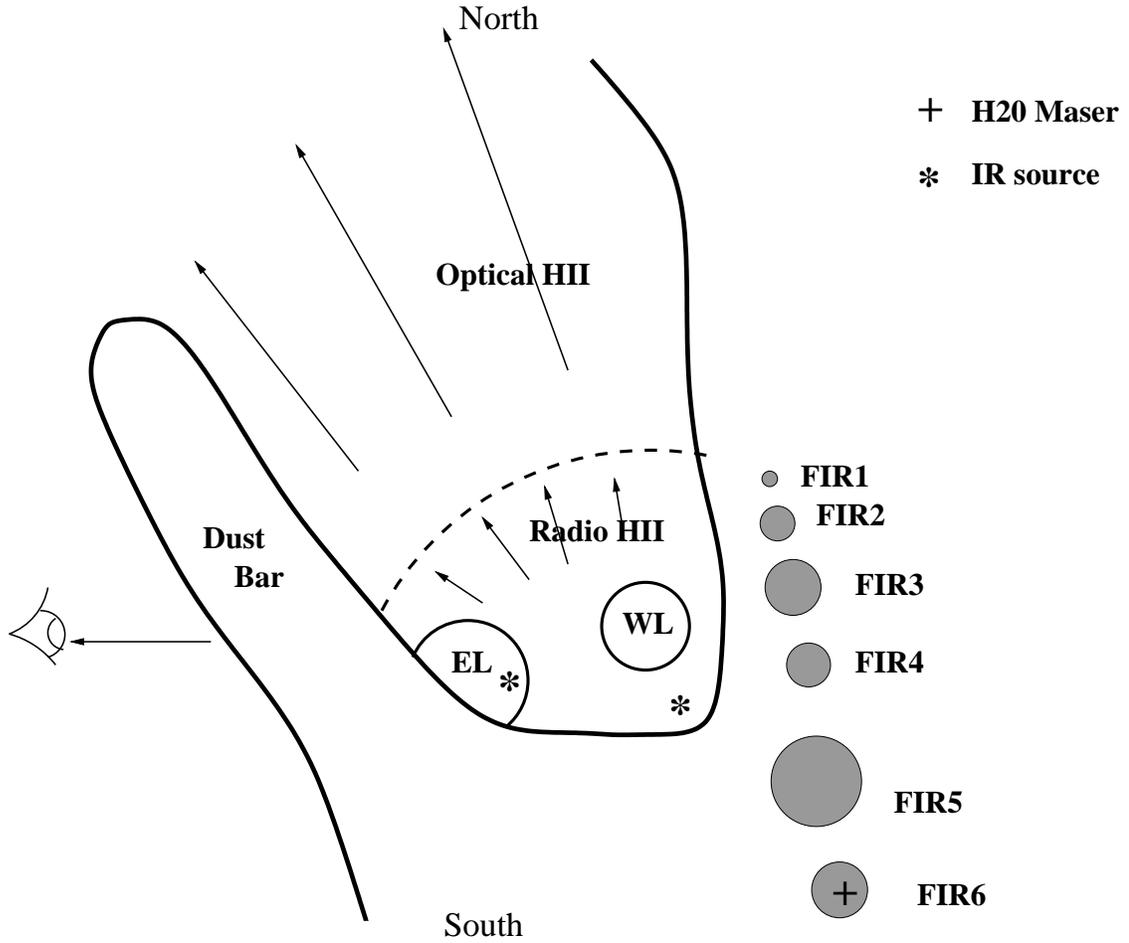}
\caption{Schematic Geometry of the NGC 2024 region.  
The geometry of the NGC 2024 \hii\, dense cores and foreground dark
dust lane, as seen from the western edge of the cloud.  From the
Earth's perspective, the dark lane (unassociated with the dense cores)
is seen in the foreground of the \hii\ region.  The \hii\ region has
blistered out of the molecular cloud to the north, west and east but
is hindered to the south by dense molecular material as indicated by
radio maps of the ionized emission
\citep{bar89}.  The IR sources marked include IRS 2 (within the
Eastern Loop) and IRS 3, near the southern boundary of the \hii\
region.}
\label{p3:mycartoon1}
\end{figure}

\begin{figure}
\vspace*{8cm}
\includegraphics{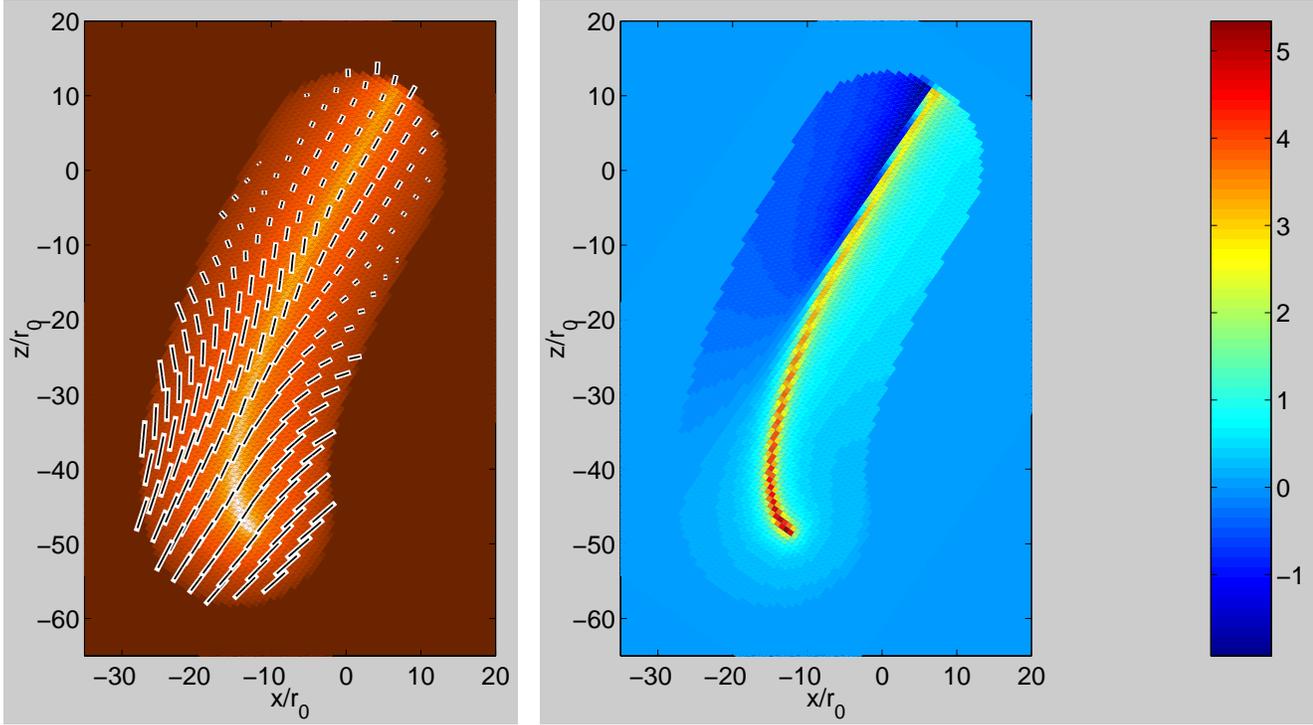}
\caption{A Helical Field Model for NGC 2024.  We show an
example of a ``bent filament'' model (discussed in Section
\ref{p3:sec:NGC2024}) for comparison with NGC 2024.  The model is a
\citet{fp00a} filament model constructed with the following
parameters: $C=1.1$, $\Gamma_z=8$, $\Gamma_\phi=15$.  The length of
the filament is $6\times10^C$ and the ends have been rounded.  We have
bent the entire filament into a circular arc perpendicular to the
plane of the sky and toward the observer, keeping the top of the
filament parallel to the original orientation.  The same pattern would
be produced for a filament bent away from the observer.  The radius of
the arc is $3/\pi$ times the filament length.  We then rotated the
entire structure by $20^\circ$ and inclined it relative to the plane
of the sky by $-15^\circ$.  The right panel shows the expected
line-of-sight field pattern from a helical field.  A reversal in field
direction is expected across the filament. }
\label{p3:model}
\end{figure}

\begin{figure}
\vspace*{15cm}
\includegraphics{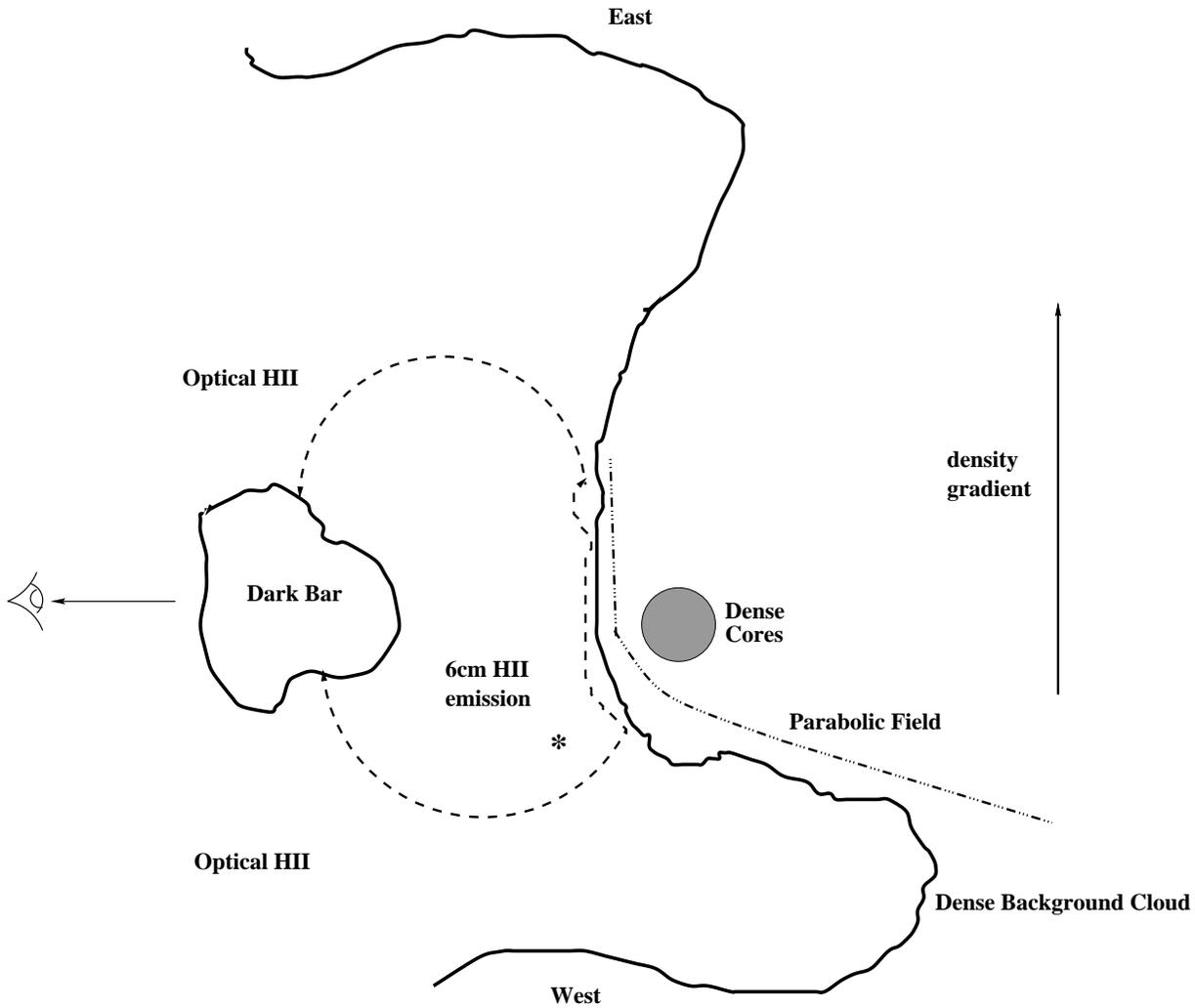}
\caption{Geometry of the NGC 2024 region, viewed from the
north.  The geometry of the NGC 2024 region as viewed from the north,
looking down on the ridge of dense cores.  A proposed orientation of
the field line is shown by the dot-dashed line.  We speculate that the
field has been compressed into a uniform layer on the edge of the
expanding \hii\ region.  The ionization front has penetrated the
molecular material more deeply to the west where the cloud is less
dense, but is hindered by the dense ridge.  This accounts for the
field lying predominantly in the plane of the sky to the east of the
ridge, but significantly in the line of sight to the west.}
\label{p3:mycartoon2}
\end{figure}

\begin{figure}
\vspace*{10cm}
\includegraphics{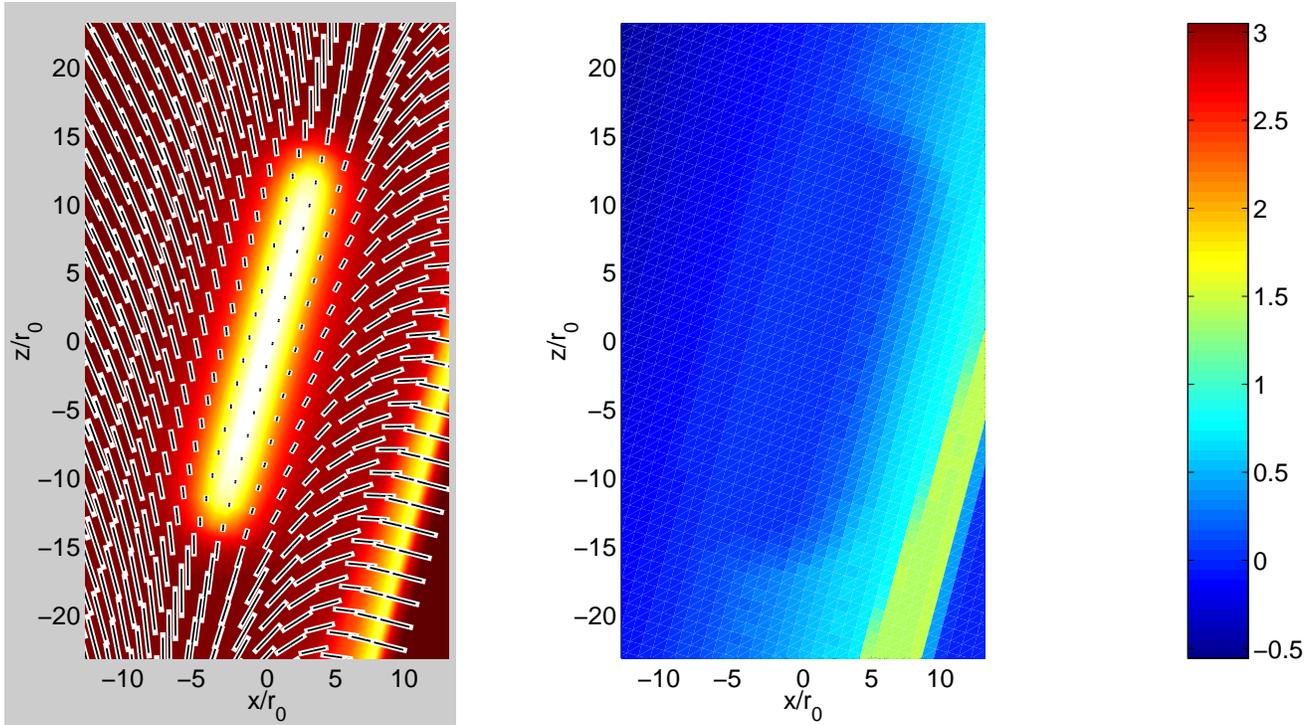}
\caption{Polarization Pattern Produced by a Parabolic
Field.  Model showing the polarization pattern from a parabolic field
bent around a dense ridge of gas, under the scenario depicted in
Figure \ref{p3:mycartoon2} and described in the text.  The left panel
shows the polarization pattern and the right traces the \blos. The
filament is unmagnetized and the field is contained entirely in the
shell of gas.  The density of the shell containing the field must be
$\sim 10\%$ that of the central axis of the filament.  The parabolic
shell is originally symmetric along the line of sight, but is rotated
by 20$^\circ$ to produce the correct behavior in the polarization and
\blos.  The inclination is 60$^\circ$ and the whole filament is
rotated by on the plane of the sky $-15^\circ$.}
\label{p3:bentfield}
\end{figure}

Tables \ref{p3:n2071table}, \ref{p3:lbs23ntable}, and \ref{p3:n2024table}
present the polarization data for the regions NGC 2071, LBS 23N, and
NGC 2024 respectively.  Each table contains only those vectors plotted
on Figures \ref{p3:n2071}, \ref{p3:lbs23}, and \ref{p3:n2024}.  Positional
offsets for NGC 2071 are from $\alpha_{J2000} = 05^{\rm h}47^{\rm
m}$5\fs16 ($05^{\rm h}44^{\rm m}$31\fs0 in B1950) and $\delta_{J2000}
= +00^\circ$ 21\arcmin\ 47\farcs1 ($+00^\circ$ 20\arcmin\ 45\farcs0 in
B1950).  The reference position for LBS 23N is $\alpha_{J2000} =
05^{\rm h}46^{\rm m}$11\fs16 ($05^{\rm h}43^{\rm m}$37\fs0 in B1950)
and $\delta_{J2000} = -00^\circ$ 09\arcmin\ 13\farcs0 ($-00^\circ$
10\arcmin\ 15\farcs2 in B1950).  Finally the reference position for
NGC 2024 is $\alpha_{J2000} = 05^{\rm h}41^{\rm m}$44\fs1 ($05^{\rm
h}39^{\rm m}$12\fs6 in B1950) and $\delta_{J2000} = -01^\circ$
54\arcmin\ 49\farcs7 ($-01^\circ$ 56\arcmin\ 15\farcs0 in B1950).

The columns in each table are: (1) the R.A. offset in arcseconds; (2)
the DEC. offset in arcseconds; (3) the polarization percentage at that
position; (4) the uncertainty in the polarization percentage; (5) the
signal-to-noise ratio of the polarization percentage; (6) the
polarization position angle; and (7) the uncertainty in the
polarization position angle.

%Table 1
\begin{deluxetable}{rrrrrrr}
\tablecolumns{7}
\tablewidth{0pc}
\tablecaption{NGC 2071 850 \micron\ Polarization Data}
\tablehead{
\colhead{$\Delta$ R.A.} & \colhead{$\Delta$ DEC.} & \colhead{$p$} & \colhead{$dp
$} & \colhead{$\sigma_p$} & \colhead{$\theta$} & \colhead{$d\theta$} \\
\colhead{(\arcsec)} & \colhead{(\arcsec)} & \colhead{(\%)} & \colhead{(\%)} & \colhead{} & \colhead{($^\circ$)} & \colhead{($^\circ$)}}
\startdata
  $  -16.5 $ &  $ -67.5 $ &   5.56 &  1.07 &   5.2 & $   63.8 $ & $  5.5 $ \\
  $   10.5 $ &  $ -58.5 $ &  12.99 &  1.41 &   9.2 & $  -14.7 $ & $  3.1 $ \\
  $    1.5 $ &  $ -58.5 $ &   8.51 &  1.33 &   6.4 & $    7.0 $ & $  4.5 $ \\
  $   -7.5 $ &  $ -58.5 $ &   5.14 &  1.04 &   4.9 & $   42.8 $ & $  5.8 $ \\
  $  -25.5 $ &  $ -58.5 $ &   6.04 &  1.40 &   4.3 & $  -47.7 $ & $  6.7 $ \\
  $  -43.5 $ &  $ -58.5 $ &   6.64 &  1.31 &   5.1 & $  -20.4 $ & $  5.7 $ \\
  $   10.5 $ &  $ -49.5 $ &  10.04 &  1.31 &   7.7 & $  -29.3 $ & $  3.7 $ \\
  $    1.5 $ &  $ -49.5 $ &   5.25 &  0.90 &   5.8 & $   18.8 $ & $  4.9 $ \\
  $  -25.5 $ &  $ -49.5 $ &   4.73 &  0.87 &   5.4 & $  -80.1 $ & $  5.3 $ \\
  $  -43.5 $ &  $ -49.5 $ &  13.76 &  1.48 &   9.3 & $   -8.0 $ & $  3.1 $ \\
  $  -52.5 $ &  $ -49.5 $ &   7.48 &  1.29 &   5.8 & $   22.0 $ & $  5.0 $ \\
  $   10.5 $ &  $ -40.5 $ &   3.54 &  0.61 &   5.9 & $  -16.1 $ & $  4.9 $ \\
  $  -16.5 $ &  $ -40.5 $ &   8.00 &  0.48 &  16.6 & $  -14.2 $ & $  1.7 $ \\
  $  -25.5 $ &  $ -40.5 $ &   4.56 &  0.68 &   6.7 & $    6.9 $ & $  4.3 $ \\
  $  -34.5 $ &  $ -40.5 $ &   9.33 &  1.46 &   6.4 & $  -19.2 $ & $  4.5 $ \\
  $  -52.5 $ &  $ -40.5 $ &  15.03 &  1.40 &  10.7 & $  -17.2 $ & $  2.7 $ \\
  $   37.5 $ &  $ -31.5 $ &   9.70 &  1.41 &   6.9 & $   46.9 $ & $  4.2 $ \\
  $   28.5 $ &  $ -31.5 $ &  11.90 &  1.02 &  11.6 & $   26.4 $ & $  2.5 $ \\
  $   19.5 $ &  $ -31.5 $ &   3.11 &  0.59 &   5.2 & $   -2.2 $ & $  5.5 $ \\
  $   10.5 $ &  $ -31.5 $ &   2.49 &  0.35 &   7.1 & $   13.0 $ & $  4.1 $ \\
  $  -16.5 $ &  $ -31.5 $ &   2.73 &  0.35 &   7.9 & $   47.0 $ & $  3.6 $ \\
  $   46.5 $ &  $ -22.5 $ &   4.34 &  1.06 &   4.1 & $   54.2 $ & $  7.0 $ \\
  $   19.5 $ &  $ -22.5 $ &   1.77 &  0.34 &   5.2 & $   -1.2 $ & $  5.5 $ \\
  $   10.5 $ &  $ -22.5 $ &   2.48 &  0.23 &  10.7 & $   26.1 $ & $  2.7 $ \\
  $    1.5 $ &  $ -22.5 $ &   1.25 &  0.26 &   4.8 & $   47.0 $ & $  6.0 $ \\
  $   -7.5 $ &  $ -22.5 $ &   1.83 &  0.21 &   8.8 & $   70.0 $ & $  3.2 $ \\
  $  -16.5 $ &  $ -22.5 $ &   1.66 &  0.25 &   6.6 & $  -85.7 $ & $  4.4 $ \\
  $  -34.5 $ &  $ -22.5 $ &   5.31 &  0.88 &   6.0 & $   21.6 $ & $  4.7 $ \\
  $  -52.5 $ &  $ -22.5 $ &   8.94 &  0.83 &  10.8 & $   49.6 $ & $  2.7 $ \\
  $   37.5 $ &  $ -13.5 $ &   5.70 &  0.93 &   6.1 & $   57.6 $ & $  4.7 $ \\
  $   28.5 $ &  $ -13.5 $ &   2.46 &  0.57 &   4.3 & $   37.1 $ & $  6.6 $ \\
  $   19.5 $ &  $ -13.5 $ &   2.29 &  0.30 &   7.6 & $   13.5 $ & $  3.8 $ \\
  $   10.5 $ &  $ -13.5 $ &   1.28 &  0.21 &   6.0 & $   18.6 $ & $  4.7 $ \\
  $    1.5 $ &  $ -13.5 $ &   1.04 &  0.17 &   6.3 & $   35.0 $ & $  4.5 $ \\
  $   -7.5 $ &  $ -13.5 $ &   1.53 &  0.18 &   8.4 & $   53.0 $ & $  3.4 $ \\
  $  -34.5 $ &  $ -13.5 $ &   3.15 &  0.31 &  10.1 & $   57.8 $ & $  2.8 $ \\
  $  -43.5 $ &  $ -13.5 $ &   5.76 &  0.50 &  11.6 & $   37.1 $ & $  2.5 $ \\
  $  -52.5 $ &  $ -13.5 $ &   4.17 &  0.68 &   6.1 & $   30.2 $ & $  4.7 $ \\
  $   37.5 $ &  $  -4.5 $ &   4.77 &  0.97 &   4.9 & $   18.0 $ & $  5.8 $ \\
  $   10.5 $ &  $  -4.5 $ &   1.74 &  0.17 &  10.3 & $   14.1 $ & $  2.8 $ \\
  $    1.5 $ &  $  -4.5 $ &   1.09 &  0.10 &  11.2 & $   29.7 $ & $  2.5 $ \\
  $   -7.5 $ &  $  -4.5 $ &   1.14 &  0.09 &  13.3 & $   55.8 $ & $  2.2 $ \\
  $  -25.5 $ &  $  -4.5 $ &   1.20 &  0.19 &   6.2 & $   55.5 $ & $  4.6 $ \\
  $  -34.5 $ &  $  -4.5 $ &   2.64 &  0.26 &  10.2 & $   75.3 $ & $  2.8 $ \\
  $  -43.5 $ &  $  -4.5 $ &   6.45 &  0.60 &  10.7 & $   37.7 $ & $  2.7 $ \\
  $  -52.5 $ &  $  -4.5 $ &   7.21 &  0.69 &  10.4 & $   13.9 $ & $  2.8 $ \\
  $   37.5 $ &  $   4.5 $ &   8.08 &  0.81 &  10.0 & $   24.3 $ & $  2.9 $ \\
  $   28.5 $ &  $   4.5 $ &   2.78 &  0.39 &   7.2 & $  -20.5 $ & $  4.0 $ \\
  $   -7.5 $ &  $   4.5 $ &   1.24 &  0.15 &   8.3 & $   61.8 $ & $  3.4 $ \\
  $  -16.5 $ &  $   4.5 $ &   1.66 &  0.17 &   9.5 & $   54.2 $ & $  3.0 $ \\
  $  -25.5 $ &  $   4.5 $ &   1.44 &  0.22 &   6.7 & $   44.0 $ & $  4.3 $ \\
  $  -43.5 $ &  $   4.5 $ &   3.73 &  0.59 &   6.3 & $   37.1 $ & $  4.5 $ \\
  $   37.5 $ &  $  13.5 $ &   5.95 &  0.76 &   7.8 & $   21.6 $ & $  3.7 $ \\
  $   28.5 $ &  $  13.5 $ &   1.43 &  0.33 &   4.3 & $  -15.0 $ & $  6.7 $ \\
  $   19.5 $ &  $  13.5 $ &   3.03 &  0.25 &  12.4 & $   -2.0 $ & $  2.3 $ \\
  $   10.5 $ &  $  13.5 $ &   1.31 &  0.25 &   5.3 & $   -2.7 $ & $  5.4 $ \\
  $   -7.5 $ &  $  13.5 $ &   1.41 &  0.24 &   5.9 & $   50.4 $ & $  4.9 $ \\
  $  -16.5 $ &  $  13.5 $ &   1.47 &  0.20 &   7.3 & $   58.9 $ & $  3.9 $ \\
  $  -25.5 $ &  $  13.5 $ &   2.18 &  0.32 &   6.8 & $   38.0 $ & $  4.2 $ \\
  $  -34.5 $ &  $  13.5 $ &   5.23 &  0.44 &  12.0 & $   21.3 $ & $  2.4 $ \\
  $  -43.5 $ &  $  13.5 $ &   4.95 &  0.57 &   8.6 & $   19.3 $ & $  3.3 $ \\
  $  -52.5 $ &  $  13.5 $ &   5.38 &  0.84 &   6.4 & $   16.8 $ & $  4.5 $ \\
  $   46.5 $ &  $  22.5 $ &   8.81 &  1.23 &   7.2 & $   28.4 $ & $  4.0 $ \\
  $   37.5 $ &  $  22.5 $ &   3.78 &  0.65 &   5.8 & $   25.3 $ & $  4.9 $ \\
  $   19.5 $ &  $  22.5 $ &   2.73 &  0.29 &   9.3 & $   29.7 $ & $  3.1 $ \\
  $   10.5 $ &  $  22.5 $ &   1.31 &  0.29 &   4.5 & $   26.9 $ & $  6.4 $ \\
  $    1.5 $ &  $  22.5 $ &   2.42 &  0.32 &   7.4 & $   74.1 $ & $  3.8 $ \\
  $   -7.5 $ &  $  22.5 $ &   1.64 &  0.32 &   5.0 & $   60.2 $ & $  5.7 $ \\
  $  -16.5 $ &  $  22.5 $ &   2.35 &  0.33 &   7.1 & $   61.0 $ & $  4.0 $ \\
  $  -25.5 $ &  $  22.5 $ &   2.60 &  0.34 &   7.6 & $   31.4 $ & $  3.8 $ \\
  $  -34.5 $ &  $  22.5 $ &   3.45 &  0.55 &   6.3 & $   21.2 $ & $  4.5 $ \\
  $  -43.5 $ &  $  22.5 $ &   5.42 &  0.54 &  10.0 & $   17.7 $ & $  2.9 $ \\
  $  -52.5 $ &  $  22.5 $ &   9.78 &  0.68 &  14.4 & $   17.8 $ & $  2.0 $ \\
  $   46.5 $ &  $  31.5 $ &  11.95 &  1.13 &  10.6 & $   27.9 $ & $  2.7 $ \\
  $   37.5 $ &  $  31.5 $ &   6.91 &  0.72 &   9.6 & $   26.4 $ & $  3.0 $ \\
  $   19.5 $ &  $  31.5 $ &   4.49 &  0.86 &   5.2 & $   52.1 $ & $  5.5 $ \\
  $   10.5 $ &  $  31.5 $ &   4.60 &  0.80 &   5.8 & $   17.6 $ & $  5.0 $ \\
  $  -25.5 $ &  $  31.5 $ &   2.74 &  0.64 &   4.3 & $   -0.8 $ & $  6.7 $ \\
  $  -34.5 $ &  $  31.5 $ &   4.45 &  0.51 &   8.8 & $   13.6 $ & $  3.3 $ \\
  $  -43.5 $ &  $  31.5 $ &   6.17 &  0.83 &   7.5 & $   -7.8 $ & $  3.8 $ \\
  $   37.5 $ &  $  40.5 $ &   5.39 &  0.73 &   7.4 & $  -84.2 $ & $  3.9 $ \\
  $   19.5 $ &  $  40.5 $ &   4.60 &  0.83 &   5.5 & $    0.8 $ & $  5.2 $ \\
  $    1.5 $ &  $  40.5 $ &   6.54 &  0.79 &   8.2 & $   14.3 $ & $  3.5 $ \\
  $  -34.5 $ &  $  40.5 $ &   3.77 &  0.78 &   4.8 & $   43.7 $ & $  5.9 $ \\
  $  -43.5 $ &  $  40.5 $ &   9.15 &  0.97 &   9.4 & $   17.7 $ & $  3.0 $ \\
  $    1.5 $ &  $  49.5 $ &   8.38 &  1.12 &   7.5 & $   17.0 $ & $  3.8 $ \\
  $   -7.5 $ &  $  49.5 $ &   7.30 &  1.04 &   7.0 & $   14.0 $ & $  4.1 $ \\
  $  -16.5 $ &  $  49.5 $ &   9.14 &  0.96 &   9.5 & $   50.5 $ & $  3.0 $ \\
  $  -34.5 $ &  $  49.5 $ &   6.52 &  0.84 &   7.8 & $   80.0 $ & $  3.7 $ \\
  $   37.5 $ &  $  58.5 $ &   3.30 &  0.80 &   4.1 & $  -38.6 $ & $  7.0 $ \\
  $   28.5 $ &  $  58.5 $ &   5.63 &  0.71 &   7.9 & $   15.2 $ & $  3.6 $ \\
  $   19.5 $ &  $  58.5 $ &   4.63 &  0.73 &   6.3 & $   25.0 $ & $  4.5 $ \\
  $    1.5 $ &  $  58.5 $ &   6.19 &  1.22 &   5.1 & $   18.3 $ & $  5.7 $ \\
  $   -7.5 $ &  $  58.5 $ &   5.60 &  1.31 &   4.3 & $    8.9 $ & $  6.7 $ \\
  $  -16.5 $ &  $  58.5 $ &   8.65 &  1.01 &   8.5 & $   19.5 $ & $  3.4 $ \\
  $  -25.5 $ &  $  58.5 $ &   3.16 &  0.69 &   4.5 & $   35.9 $ & $  6.3 $ \\
  $  -34.5 $ &  $  58.5 $ &   3.84 &  0.91 &   4.2 & $   33.1 $ & $  6.8 $ \\
  $   19.5 $ &  $  67.5 $ &  14.62 &  1.33 &  11.0 & $   64.9 $ & $  2.6 $ \\
  $   10.5 $ &  $  67.5 $ &  10.32 &  1.19 &   8.7 & $   36.5 $ & $  3.3 $ \\
\enddata
\label{p3:n2071table}
\end{deluxetable}

%Table 2
\begin{deluxetable}{rrrrrrr}
\tablecolumns{7}
\tablewidth{0pc}
\tablecaption{LBS 23N 850 \micron\ Polarization Data}
\tablehead{
\colhead{$\Delta$ R.A.} & \colhead{$\Delta$ DEC.} & \colhead{$p$} & \colhead{$dp
$} & \colhead{$\sigma_p$} & \colhead{$\theta$} & \colhead{$d\theta$} \\
\colhead{(\arcsec)} & \colhead{(\arcsec)} & \colhead{(\%)} & \colhead{(\%)} & \colhead{} & \colhead{($^\circ$)} & \colhead{($^\circ$)}}
\startdata
  $   -9.0 $ &  $-124.5 $ &   5.58 &  1.03 &   5.4 &  $ -29.1 $ &  $ 5.3 $ \\
  $   15.0 $ &  $-112.5 $ &  12.03 &  1.37 &   8.8 &  $  29.9 $ &  $ 3.3 $ \\
  $    3.0 $ &  $-112.5 $ &   4.29 &  0.60 &   7.1 &  $  38.4 $ &  $ 4.0 $ \\
  $   -9.0 $ &  $-112.5 $ &   1.83 &  0.30 &   6.1 &  $ -16.7 $ &  $ 4.7 $ \\
  $   15.0 $ &  $-100.5 $ &  11.75 &  1.25 &   9.4 &  $   5.1 $ &  $ 3.1 $ \\
  $    3.0 $ &  $-100.5 $ &   5.34 &  0.68 &   7.8 &  $  74.4 $ &  $ 3.7 $ \\
  $   -9.0 $ &  $-100.5 $ &   3.14 &  0.34 &   9.1 &  $ -82.0 $ &  $ 3.1 $ \\
  $  -21.0 $ &  $-100.5 $ &   2.88 &  0.30 &   9.7 &  $ -88.8 $ &  $ 3.0 $ \\
  $  -33.0 $ &  $-100.5 $ &   3.12 &  0.62 &   5.0 &  $  71.7 $ &  $ 5.7 $ \\
  $  -45.0 $ &  $-100.5 $ &   8.47 &  1.35 &   6.3 &  $  75.1 $ &  $ 4.5 $ \\
  $   15.0 $ &  $-88.5 $ &   8.27 &  1.12 &   7.4 &  $ -11.9 $ &  $ 3.9 $ \\
  $    3.0 $ &  $-88.5 $ &   4.88 &  0.68 &   7.1 &  $  -3.4 $ &  $ 4.0 $ \\
  $   -9.0 $ &  $-88.5 $ &   1.97 &  0.49 &   4.1 &  $ -88.2 $ &  $ 7.1 $ \\
  $  -21.0 $ &  $-88.5 $ &   2.94 &  0.50 &   5.9 &  $  29.4 $ &  $ 4.9 $ \\
  $  -33.0 $ &  $-88.5 $ &   3.77 &  0.76 &   5.0 &  $  71.7 $ &  $ 5.7 $ \\
  $  -45.0 $ &  $-88.5 $ &   5.28 &  1.28 &   4.1 &  $  51.7 $ &  $ 6.9 $ \\
  $   27.0 $ &  $-76.5 $ &  13.00 &  1.34 &   9.7 &  $  28.8 $ &  $ 3.0 $ \\
  $   15.0 $ &  $-76.5 $ &   5.45 &  0.80 &   6.9 &  $  21.6 $ &  $ 4.2 $ \\
  $    3.0 $ &  $-76.5 $ &   3.61 &  0.65 &   5.6 &  $ -26.6 $ &  $ 5.1 $ \\
  $   -9.0 $ &  $-76.5 $ &   6.55 &  0.57 &  11.5 &  $ -53.6 $ &  $ 2.5 $ \\
  $  -21.0 $ &  $-76.5 $ &   7.21 &  0.77 &   9.4 &  $ -42.3 $ &  $ 3.1 $ \\
  $  -45.0 $ &  $-76.5 $ &  12.16 &  1.20 &  10.2 &  $   7.8 $ &  $ 2.8 $ \\
  $   27.0 $ &  $-64.5 $ &   7.82 &  1.29 &   6.0 &  $   1.8 $ &  $ 4.7 $ \\
  $   15.0 $ &  $-64.5 $ &   3.21 &  0.63 &   5.1 &  $  10.3 $ &  $ 5.6 $ \\
  $   -9.0 $ &  $-64.5 $ &   2.52 &  0.47 &   5.4 &  $ -78.8 $ &  $ 5.4 $ \\
  $  -21.0 $ &  $-64.5 $ &   3.40 &  0.71 &   4.8 &  $  62.8 $ &  $ 5.9 $ \\
  $   27.0 $ &  $-52.5 $ &  12.61 &  1.01 &  12.5 &  $   8.1 $ &  $ 2.3 $ \\
  $   15.0 $ &  $-52.5 $ &   6.77 &  0.47 &  14.5 &  $   4.9 $ &  $ 2.0 $ \\
  $    3.0 $ &  $-52.5 $ &   1.47 &  0.26 &   5.7 &  $  16.3 $ &  $ 5.0 $ \\
  $   -9.0 $ &  $-52.5 $ &   2.61 &  0.23 &  11.5 &  $   5.2 $ &  $ 2.5 $ \\
  $  -21.0 $ &  $-52.5 $ &   2.09 &  0.42 &   5.0 &  $  50.1 $ &  $ 5.7 $ \\
  $  -33.0 $ &  $-52.5 $ &   3.51 &  0.67 &   5.3 &  $   0.6 $ &  $ 5.4 $ \\
  $   27.0 $ &  $-40.5 $ &   8.82 &  0.74 &  11.9 &  $   6.9 $ &  $ 2.4 $ \\
  $   15.0 $ &  $-40.5 $ &   2.61 &  0.37 &   7.1 &  $  32.3 $ &  $ 4.1 $ \\
  $    3.0 $ &  $-40.5 $ &   1.55 &  0.13 &  11.6 &  $  23.5 $ &  $ 2.5 $ \\
  $   27.0 $ &  $-28.5 $ &   6.18 &  1.06 &   5.8 &  $  38.5 $ &  $ 4.9 $ \\
  $   15.0 $ &  $-28.5 $ &   5.57 &  0.41 &  13.5 &  $  28.3 $ &  $ 2.1 $ \\
  $    3.0 $ &  $-28.5 $ &   1.58 &  0.18 &   8.8 &  $  27.8 $ &  $ 3.2 $ \\
  $   -9.0 $ &  $-28.5 $ &   1.40 &  0.13 &  10.9 &  $  46.0 $ &  $ 2.6 $ \\
  $  -21.0 $ &  $-28.5 $ &   1.98 &  0.24 &   8.4 &  $   7.9 $ &  $ 3.4 $ \\
  $  -33.0 $ &  $-28.5 $ &   5.60 &  0.63 &   8.9 &  $  -6.9 $ &  $ 3.2 $ \\
  $   51.0 $ &  $-16.5 $ &   8.44 &  1.14 &   7.4 &  $  17.0 $ &  $ 3.9 $ \\
  $   15.0 $ &  $-16.5 $ &   7.53 &  0.92 &   8.2 &  $  29.0 $ &  $ 3.5 $ \\
  $    3.0 $ &  $-16.5 $ &   2.23 &  0.41 &   5.5 &  $  11.5 $ &  $ 5.2 $ \\
  $   -9.0 $ &  $-16.5 $ &   1.47 &  0.30 &   4.9 &  $  36.3 $ &  $ 5.9 $ \\
  $  -21.0 $ &  $-16.5 $ &   2.83 &  0.36 &   7.9 &  $  15.6 $ &  $ 3.6 $ \\
  $  -45.0 $ &  $-16.5 $ &  12.59 &  1.23 &  10.2 &  $  30.8 $ &  $ 2.8 $ \\
  $   39.0 $ &  $-4.5 $ &   6.46 &  1.17 &   5.5 &  $  -4.9 $ &  $ 5.2 $ \\
  $   15.0 $ &  $-4.5 $ &  21.96 &  1.30 &  16.8 &  $  -4.4 $ &  $ 1.7 $ \\
  $    3.0 $ &  $-4.5 $ &   2.54 &  0.54 &   4.7 &  $  15.1 $ &  $ 6.1 $ \\
  $  -33.0 $ &  $-4.5 $ &   4.97 &  0.58 &   8.6 &  $ -29.4 $ &  $ 3.3 $ \\
  $   39.0 $ &  $7.5 $ &   4.09 &  0.77 &   5.3 &  $  29.8 $ &  $ 5.4 $ \\
  $   27.0 $ &  $7.5 $ &   8.46 &  1.11 &   7.7 &  $  19.7 $ &  $ 3.7 $ \\
  $   15.0 $ &  $7.5 $ &   6.71 &  0.83 &   8.1 &  $  16.4 $ &  $ 3.6 $ \\
  $    3.0 $ &  $7.5 $ &   4.21 &  0.53 &   8.0 &  $ -28.8 $ &  $ 3.6 $ \\
  $   -9.0 $ &  $7.5 $ &   1.17 &  0.23 &   5.0 &  $  45.7 $ &  $ 5.8 $ \\
  $  -21.0 $ &  $7.5 $ &   1.26 &  0.28 &   4.5 &  $ -10.5 $ &  $ 6.3 $ \\
  $  -33.0 $ &  $7.5 $ &   3.28 &  0.50 &   6.5 &  $  20.0 $ &  $ 4.4 $ \\
  $   39.0 $ &  $19.5 $ &  11.24 &  1.09 &  10.3 &  $ -62.0 $ &  $ 2.8 $ \\
  $   27.0 $ &  $19.5 $ &  10.48 &  1.05 &   9.9 &  $ -30.8 $ &  $ 2.9 $ \\
  $   15.0 $ &  $19.5 $ &   5.70 &  0.58 &   9.8 &  $  21.4 $ &  $ 2.9 $ \\
  $    3.0 $ &  $19.5 $ &   4.01 &  0.44 &   9.2 &  $ -23.6 $ &  $ 3.1 $ \\
  $   -9.0 $ &  $19.5 $ &   5.57 &  0.43 &  12.9 &  $  14.6 $ &  $ 2.2 $ \\
  $  -21.0 $ &  $19.5 $ &   2.68 &  0.44 &   6.1 &  $  14.9 $ &  $ 4.7 $ \\
  $  -33.0 $ &  $19.5 $ &   7.10 &  0.43 &  16.4 &  $  -6.7 $ &  $ 1.8 $ \\
  $  -45.0 $ &  $19.5 $ &   1.89 &  0.46 &   4.1 &  $  26.3 $ &  $ 7.0 $ \\
  $    3.0 $ &  $31.5 $ &   5.46 &  0.66 &   8.2 &  $  57.2 $ &  $ 3.5 $ \\
  $  -21.0 $ &  $31.5 $ &  11.41 &  0.66 &  17.3 &  $  -4.8 $ &  $ 1.7 $ \\
  $  -33.0 $ &  $31.5 $ &   5.90 &  0.57 &  10.3 &  $  36.8 $ &  $ 2.8 $ \\
  $  -45.0 $ &  $31.5 $ &   7.10 &  0.78 &   9.1 &  $  11.6 $ &  $ 3.2 $ \\
\enddata
\label{p3:lbs23ntable}
\end{deluxetable}

%Table 3
\begin{deluxetable}{rrrrrrr}
\tablecolumns{7} 
\tablewidth{0pc} 
\tablecaption{NGC 2024 850 \micron\ Polarization Data} 
\tablehead{ \colhead{$\Delta$ R.A.} &
\colhead{$\Delta$ DEC.} & \colhead{$p$} & \colhead{$dp $} &
\colhead{$\sigma_p$} & \colhead{$\theta$} & \colhead{$d\theta$} \\
\colhead{(\arcsec)} & \colhead{(\arcsec)} & \colhead{(\%)} &
\colhead{(\%)} & \colhead{} & \colhead{($^\circ$)} &
\colhead{($^\circ$)}} 
\startdata 
$ 13.5$ & $-147.0 $ & 6.77 & 1.13 & 6.0 & $ -31.0$ & $ 4.8 $ \\ 
$ 1.5$ & $-147.0 $ & 4.99 & 0.90 & 5.5 & $ -17.9$ & $ 5.2 $ \\ 
$ 49.5$ & $-135.0 $ & 9.39 & 0.93 & 10.1 & $ -18.8$ & $ 2.8 $ \\ 
$ 13.5$ & $-135.0 $ & 4.28 & 0.38 & 11.3 & $ -41.6$ & $ 2.5 $ \\ 
$ 1.5$ & $-135.0 $ & 5.81 & 0.32 & 18.2 & $ -38.1$ & $ 1.6 $ \\ 
$ -10.5$ & $-135.0 $ & 7.63 & 0.44 & 17.2 & $ -24.5$ & $1.7 $ \\ 
$ -22.5$ & $-135.0 $ & 8.92 & 0.90 & 9.9 & $ 51.7$ & $ 2.9 $ \\ 
$ 49.5$ & $-123.0 $ & 4.96 & 0.64 & 7.7 & $ 3.3$ & $ 3.7 $ \\ 
$ 37.5$ & $-123.0 $ & 5.35 & 0.60 & 8.9 & $ -22.2$ & $ 3.2 $ \\ 
$ 25.5$ & $-123.0 $ & 1.87 & 0.32 & 5.8 & $ -29.2$ & $ 4.9 $ \\ 
$ 13.5$ & $-123.0 $ & 4.50 & 0.15 & 29.7 & $ -39.0$ & $ 1.0 $ \\ 
$ 1.5$ & $-123.0 $ & 4.03 & 0.27 & 15.2 & $ -59.8$ & $ 1.9 $ \\ 
$ -10.5$ & $-123.0 $ & 4.76 & 0.39 & 12.2 & $ -46.0$ & $ 2.3 $ \\ 
$ -22.5$ & $-123.0 $ & 3.95 & 0.50 & 8.0 & $ -47.0$ & $ 3.6 $ \\ 
$ 37.5$ & $-111.0 $ & 3.12 & 0.37 & 8.5 & $ -11.0$ & $ 3.4 $ \\ 
$ 13.5$ & $-111.0 $ & 1.94 & 0.19 & 10.2 & $ -50.5$ & $ 2.8 $ \\ 
$ 1.5$ & $-111.0 $ & 2.05 & 0.30 & 6.8 & $ -67.8$ & $ 4.2 $ \\ 
$ -10.5$ & $-111.0 $ & 6.07 & 0.35 & 17.1 & $ -33.0$ & $ 1.7 $ \\ 
$ -22.5$ & $-111.0 $ & 3.07 & 0.42 & 7.3 & $ -38.7$ & $ 3.9 $ \\ 
$ 61.5$ & $ -99.0 $ & 9.10 & 0.90 & 10.1 & $ -13.4$ & $ 2.8 $ \\ 
$ 49.5$ & $ -99.0 $ & 4.27 & 0.37 & 11.6 & $ -9.6$ & $ 2.5 $ \\ 
$ 37.5$ & $ -99.0 $ & 1.37 & 0.26 & 5.4 & $ 23.6$ & $ 5.4 $ \\ 
$ 13.5$ & $ -99.0 $ & 2.11 & 0.13 & 16.0 & $ -35.9$ & $ 1.8 $ \\ 
$ 1.5$ & $ -99.0 $ & 2.22 & 0.23 & 9.7 & $ -41.3$ & $ 3.0 $ \\ 
$ -10.5$ & $ -99.0 $ & 1.96 & 0.29 & 6.8 & $ -39.4$ & $ 4.2 $ \\ 
$ -34.5$ & $ -99.0 $ & 3.57 & 0.35 & 10.1 & $ -60.2$ & $ 2.8 $ \\ 
$ -46.5$ & $ -99.0 $ & 2.18 & 0.47 & 4.7 & $ 81.4$ & $ 6.1 $ \\ 
$ -58.5$ & $ -99.0 $ & 5.21 & 1.08 & 4.8 & $ -10.8$ & $ 5.9 $ \\ 
$ 61.5$ & $ -87.0 $ & 3.55 & 0.42 & 8.4 & $ 9.3$ & $ 3.4 $ \\
$ 49.5$ & $ -87.0 $ & 2.14 & 0.25 & 8.7 & $ -8.7$ & $ 3.3 $ \\ 
$ 37.5$ & $ -87.0 $ & 2.21 & 0.20 & 11.3 & $ 35.7$ & $ 2.5 $ \\ 
$ 25.5$ & $ -87.0 $ & 1.11 & 0.15 & 7.5 & $ 33.0$ & $ 3.8 $ \\ 
$ 13.5$ & $ -87.0 $ & 1.12 & 0.10 & 11.2 & $ -33.3$ & $ 2.6 $ \\ 
$ 1.5$ & $ -87.0 $ & 2.65 & 0.17 & 15.9 & $ -27.0$ & $ 1.8 $ \\ 
$ -10.5$ & $ -87.0 $ & 2.38 & 0.19 & 12.3 & $ -16.2$ & $ 2.3 $ \\ 
$ -22.5$ & $ -87.0 $ & 2.02 & 0.19 & 10.7 & $ -24.0$ & $ 2.7 $ \\ 
$ -34.5$ & $ -87.0 $ & 1.47 & 0.33 & 4.5 & $ -65.9$ & $ 6.4 $ \\ 
$ -46.5$ & $ -87.0 $ & 5.39 & 0.60 & 9.0 & $ 74.8$ & $ 3.2 $ \\ 
$ 73.5$ & $ -75.0 $ & 7.90 & 0.98 & 8.0 & $ -3.6$ & $ 3.6 $ \\ 
$ 61.5$ & $ -75.0 $ & 4.94 & 0.35 & 13.9 & $ -4.9$ & $ 2.1 $ \\ 
$ 49.5$ & $ -75.0 $ & 2.70 & 0.19 & 14.4 & $ 10.1$ & $ 2.0 $ \\ 
$ 37.5$ & $ -75.0 $ & 1.41 & 0.17 & 8.2 & $ 17.2$ & $ 3.5 $ \\ 
$ 1.5$ & $ -75.0 $ & 1.31 & 0.12 & 10.5 & $ -23.7$ & $ 2.7 $ \\ 
$ -10.5$ & $ -75.0 $ & 1.63 & 0.15 & 10.7 & $ -20.7$ & $ 2.7 $ \\ 
$ -22.5$ & $ -75.0 $ & 1.77 & 0.16 & 10.8 & $ -26.4$ & $ 2.7 $ \\ 
$ -34.5$ & $ -75.0 $ & 2.81 & 0.25 & 11.3 & $ -10.2$ & $ 2.5 $ \\ 
$ -46.5$ & $ -75.0 $ & 3.50 & 0.52 & 6.8 & $ -22.0$ & $ 4.2 $ \\ 
$ 73.5$ & $ -63.0 $ & 5.45 & 0.99 & 5.5 & $ -28.7$ & $ 5.2 $ \\ 
$ 61.5$ & $ -63.0 $ & 4.26 & 0.35 & 12.3 & $ 5.7$ & $ 2.3 $ \\ 
$ 49.5$ & $ -63.0 $ & 2.73 & 0.16 & 16.9 & $ 24.9$ & $ 1.7 $ \\ 
$ 37.5$ & $ -63.0 $ & 1.38 & 0.14 & 10.0 & $ 7.6$ & $ 2.9 $ \\ 
$ 1.5$ & $ -63.0 $ & 2.01 & 0.09 & 21.9 & $ -48.0$ & $ 1.3 $ \\ 
$ -10.5$ & $ -63.0 $ & 1.27 & 0.12 & 10.6 & $ -41.8$ & $ 2.7 $ \\ 
$ -22.5$ & $ -63.0 $ & 1.67 & 0.13 & 13.0 & $ -19.0$ & $ 2.2 $ \\ 
$ -34.5$ & $ -63.0 $ & 3.24 & 0.21 & 15.7 & $ -32.1$ & $ 1.8 $ \\ 
$ -46.5$ & $ -63.0 $ & 11.87 & 0.57 & 20.8 & $ -43.6$ & $ 1.4 $ \\ 
$ 73.5$ & $ -51.0 $ & 4.43 & 0.55 & 8.1 & $ -7.3$ & $ 3.5 $ \\ 
$ 61.5$ & $ -51.0 $ & 3.65 & 0.25 & 14.7 & $ -0.6$ & $ 2.0 $ \\ 
$ 49.5$ & $ -51.0 $ & 2.30 & 0.14 & 16.3 & $ 14.6$ & $ 1.8 $ \\ 
$ 37.5$ & $ -51.0 $ & 1.33 & 0.09 & 14.0 & $ 20.9$ & $ 2.0 $ \\ 
$ 25.5$ & $ -51.0 $ & 1.23 & 0.09 & 14.0 & $ 11.9$ & $ 2.0 $ \\ 
$ 1.5$ & $ -51.0 $ & 1.42 & 0.04 & 32.4 & $ -48.3$ & $ 0.9 $ \\ 
$ -22.5$ & $ -51.0 $ & 1.84 & 0.13 & 14.0 & $ -52.1$ & $ 2.0 $ \\ 
$ -34.5$ & $ -51.0 $ & 3.01 & 0.27 & 11.2 & $ -47.6$ & $ 2.6 $ \\ 
$ -46.5$ & $ -51.0 $ & 10.69 & 0.86 & 12.4 & $ -60.8$ & $ 2.3 $ \\ 
$ 73.5$ & $ -39.0 $ & 9.98 & 0.69 & 14.4 & $ 6.6$ & $ 2.0 $ \\ 
$ 61.5$ & $ -39.0 $ & 3.22 & 0.24 & 13.3 & $ 3.7$ & $ 2.2 $ \\ 
$ 49.5$ & $ -39.0 $ & 1.81 & 0.13 & 14.2 & $ 21.7$ & $ 2.0 $ \\ 
$ 37.5$ & $ -39.0 $ & 1.20 & 0.10 & 11.5 & $ 34.7$ & $ 2.5 $ \\ 
$ 25.5$ & $ -39.0 $ & 1.61 & 0.10 & 16.5 & $ 37.9$ & $ 1.7 $ \\ 
$ 1.5$ & $ -39.0 $ & 1.02 & 0.09 & 11.8 & $ -39.7$ & $ 2.4 $ \\ 
$ -10.5$ & $ -39.0 $ & 1.45 & 0.09 & 16.0 & $ -50.1$ & $ 1.8 $ \\ 
$ -22.5$ & $ -39.0 $ & 2.14 & 0.19 & 11.6 & $ -74.0$ & $ 2.5 $ \\ 
$ -34.5$ & $ -39.0 $ & 4.56 & 0.36 & 12.8 & $ -74.1$ & $ 2.2 $ \\ 
$ -46.5$ & $ -39.0 $ & 6.39 & 0.63 & 10.2 & $ -74.0$ & $ 2.8 $ \\ 
$ 61.5$ & $ -27.0 $ & 2.27 & 0.30 & 7.5 & $ 3.5$ & $ 3.8 $ \\ 
$ 49.5$ & $ -27.0 $ & 1.95 & 0.16 & 12.1 & $ 25.7$ & $ 2.4$ \\ 
$ 37.5$ & $ -27.0 $ & 2.24 & 0.14 & 16.5 & $ 34.0$ & $ 1.7 $ \\ 
$ 25.5$ & $ -27.0 $ & 2.34 & 0.13 & 17.4 & $ 44.3$ & $ 1.6 $ \\ 
$ -10.5$ & $ -27.0 $ & 1.29 & 0.10 & 12.8 & $ -72.7$ & $ 2.2 $ \\ 
$ -22.5$ & $ -27.0 $ & 2.53 & 0.18 & 13.9 & $ -66.6$ & $ 2.1 $ \\ 
$ -34.5$ & $ -27.0 $ & 4.69 & 0.31 & 15.1 & $ -80.7$ & $ 1.9 $ \\ 
$ -46.5$ & $ -27.0 $ & 4.83 & 0.35 & 13.9 & $ -74.6$ & $ 2.1 $ \\ 
$ -58.5$ & $ -27.0 $ & 2.30 & 0.49 & 4.7 & $ -47.5$ & $ 6.1 $ \\ 
$ 73.5$ & $ -15.0 $ & 7.29 & 0.92 & 7.9 & $ 13.9$ & $ 3.6 $ \\ 
$ 49.5$ & $ -15.0 $ & 1.39 & 0.19 & 7.3 & $ 25.1$ & $ 3.9 $ \\ 
$ 37.5$ & $ -15.0 $ & 1.75 & 0.13 & 13.3 & $ 38.3$ & $ 2.2 $ \\ 
$ 25.5$ & $ -15.0 $ & 1.93 & 0.14 & 14.1 & $ 36.9$ & $ 2.0 $ \\ 
$ -22.5$ & $ -15.0 $ & 1.45 & 0.16 & 9.1 & $ -67.7$ & $ 3.2 $ \\ 
$ -34.5$ & $ -15.0 $ & 3.52 & 0.29 & 12.1 & $ -67.8$ & $ 2.4 $ \\ 
$ -46.5$ & $ -15.0 $ & 1.35 & 0.28 & 4.9 & $ -82.1$ & $ 5.9 $ \\ 
$ -58.5$ & $ -15.0 $ & 3.81 & 0.50 & 7.6 & $ -85.3$ & $ 3.8 $ \\ 
$ 73.5$ & $ -3.0 $ & 6.76 & 0.96 & 7.0 & $ 29.1$ & $ 4.1 $ \\ 
$ 61.5$ & $ -3.0 $ & 4.65 & 0.47 & 10.0 & $ 34.9$ & $ 2.9 $ \\ 
$ 49.5$ & $ -3.0 $ & 2.66 & 0.21 & 12.5 & $ 19.0$ & $ 2.3 $ \\ 
$ 37.5$ & $ -3.0 $ & 2.44 & 0.15 & 16.8 & $ 37.7$ & $ 1.7 $ \\ 
$ 25.5$ & $ -3.0 $ & 1.32 & 0.12 & 11.1 & $ 31.6$ & $ 2.6 $ \\ 
$ -10.5$ & $ -3.0 $ & 1.05 & 0.10 & 10.3 & $ -61.8$ & $ 2.8 $ \\ 
$ -22.5$ & $ -3.0 $ & 2.08 & 0.13 & 16.5 & $ -62.1$ & $ 1.7 $ \\ 
$ -34.5$ & $ -3.0 $ & 1.86 & 0.22 & 8.5 & $ -84.5$ & $ 3.4 $ \\ 
$ -46.5$ & $ -3.0 $ & 3.92 & 0.33 & 12.1 & $ -82.4$ & $ 2.4 $ \\ 
$ 73.5$ & $ 9.0 $ & 13.41 & 1.13 & 11.9 & $ 37.8$ & $ 2.4 $ \\ 
$ 61.5$ & $ 9.0 $ & 5.49 & 0.58 & 9.5 & $ 26.8$ & $ 3.0 $ \\ 
$ 49.5$ & $ 9.0 $ & 2.86 & 0.24 & 11.8 & $ 16.8$ & $ 2.4 $ \\ 
$ 37.5$ & $ 9.0 $ & 1.51 & 0.16 & 9.2 & $ 22.4$ & $ 3.1 $ \\ 
$ 25.5$ & $ 9.0 $ & 2.77 & 0.11 & 25.1 & $ 1.6$ & $ 1.1 $ \\ 
$ 13.5$ & $ 9.0 $ & 2.13 & 0.10 & 20.6 & $ 1.3$ & $ 1.4 $ \\ 
$ 1.5$ & $ 9.0 $ & 1.27 & 0.06 & 20.9 & $ -23.8$ & $ 1.4 $ \\ 
$ -22.5$ & $ 9.0 $ & 2.03 & 0.19 & 10.6 & $ -67.1$ & $ 2.7 $ \\ 
$ -34.5$ & $ 9.0 $ & 3.35 & 0.23 & 14.4 & $ -76.1$ & $ 2.0 $ \\ 
$ -46.5$ & $ 9.0 $ & 6.46 & 0.33 & 19.7 & $ -87.7$ & $ 1.5 $ \\ 
$ -58.5$ & $ 9.0 $ & 4.36 & 0.50 & 8.8 & $ -62.4$ & $ 3.3 $ \\ 
$ 49.5$ & $ 21.0 $ & 2.54 & 0.36 & 7.0 & $ 22.0$ & $ 4.1 $ \\ 
$ 37.5$ & $ 21.0 $ & 1.89 & 0.21 & 9.2 & $ 2.2$ & $ 3.1 $ \\ 
$ 25.5$ & $ 21.0 $ & 3.32 & 0.14 & 24.5 & $ -9.5$ & $ 1.2 $ \\ 
$ 13.5$ & $ 21.0 $ & 2.30 & 0.12 & 19.1 & $ -1.0$ & $ 1.5 $ \\ 
$ 1.5$ & $ 21.0 $ & 1.98 & 0.13 & 15.1 & $ 3.2$ & $ 1.9 $ \\ 
$ -10.5$ & $ 21.0 $ & 1.35 & 0.08 & 17.7 & $ -28.7$ & $ 1.6 $ \\ 
$ -22.5$ & $ 21.0 $ & 1.19 & 0.13 & 8.9 & $ -43.9$ & $ 3.2 $ \\ 
$ -34.5$ & $ 21.0 $ & 4.46 & 0.23 & 19.6 & $ -79.7$ & $ 1.5 $ \\ 
$ -46.5$ & $ 21.0 $ & 4.14 & 0.28 & 14.6 & $ -69.5$ & $ 2.0 $ \\ 
$ -58.5$ & $ 21.0 $ & 2.91 & 0.35 & 8.4 & $ -89.4$ & $ 3.4 $ \\ 
$ 49.5$ & $ 33.0 $ & 4.16 & 0.77 & 5.4 & $ -75.3$ & $ 5.3 $ \\ 
$ 37.5$ & $ 33.0 $ & 2.35 & 0.34 & 6.9 & $ -26.5$ & $ 4.2 $ \\ 
$ 25.5$ & $ 33.0 $ & 3.73 & 0.25 & 15.1 & $ -2.8$ & $ 1.9 $ \\ 
$ 13.5$ & $ 33.0 $ & 2.21 & 0.20 & 11.1 & $ 1.1$ & $ 2.6 $ \\ 
$ 1.5$ & $ 33.0 $ & 2.69 & 0.15 & 18.1 & $ 4.1$ & $ 1.6 $ \\ 
$ -10.5$ & $ 33.0 $ & 1.63 & 0.17 & 9.9 & $ -19.7$ & $ 2.9 $ \\ 
$ -22.5$ & $ 33.0 $ & 1.80 & 0.07 & 25.5 & $ -35.0$ & $ 1.1 $ \\ 
$ -34.5$ & $ 33.0 $ & 2.56 & 0.24 & 10.5 & $ -81.5$ & $ 2.7 $ \\ 
$ -46.5$ & $ 33.0 $ & 1.32 & 0.23 & 5.6 & $ -68.4$ & $ 5.1 $ \\ 
$ -58.5$ & $ 33.0 $ & 1.61 & 0.35 & 4.6 & $ -84.0$ & $ 6.2 $ \\ 
$ 37.5$ & $ 45.0 $ & 4.28 & 0.51 & 8.3 & $ -42.2$ & $ 3.4 $ \\ 
$ 25.5$ & $ 45.0 $ & 3.07 & 0.42 & 7.4 & $ -8.3$ & $ 3.9 $ \\ 
$ 13.5$ & $ 45.0 $ & 1.23 & 0.29 & 4.3 & $ 29.2$ & $ 6.7 $ \\ 
$ 1.5$ & $ 45.0 $ & 1.06 & 0.19 & 5.6 & $ -25.6$ & $ 5.2 $ \\ 
$ -10.5$ & $ 45.0 $ & 1.39 & 0.15 & 9.0 & $ -28.6$ & $ 3.2 $ \\ 
$ -22.5$ & $ 45.0 $ & 1.54 & 0.11 & 14.3 & $ -24.8$ & $ 2.0 $ \\ 
$ -34.5$ & $ 45.0 $ & 1.56 & 0.14 & 11.4 & $ -82.5$ & $ 2.5 $ \\ 
$ -58.5$ & $ 45.0 $ & 2.50 & 0.29 & 8.6 & $ 77.7$ & $ 3.3 $ \\ 
$ -70.5$ & $ 45.0 $ & 6.64 & 1.16 & 5.7 & $ 52.6$ & $ 5.0 $ \\ 
$ 25.5$ & $ 57.0 $ & 6.36 & 0.61 & 10.4 & $ -48.2$ & $ 2.7 $ \\ 
$ 13.5$ & $ 57.0 $ & 2.92 & 0.59 & 5.0 & $ 45.8$ & $ 5.8 $ \\ 
$ -22.5$ & $ 57.0 $ & 1.29 & 0.25 & 5.2 & $ -15.0$ & $ 5.5 $ \\ 
$ -34.5$ & $ 57.0 $ & 1.50 & 0.14 & 10.4 & $ -68.0$ & $ 2.8 $ \\ 
$ -46.5$ & $ 57.0 $ & 1.28 & 0.21 & 6.0 & $ -86.9$ & $ 4.8 $ \\ 
$ -58.5$ & $ 57.0 $ & 3.11 & 0.49 & 6.3 & $ 46.3$ & $ 4.5 $ \\ 
$ 49.5$ & $ 69.0 $ & 7.34 & 0.83 & 8.8 & $ 52.7$ & $ 3.3 $ \\ 
$ 37.5$ & $ 69.0 $ & 6.92 & 0.78 & 8.8 & $ 32.4$ & $ 3.2 $ \\ 
$ 25.5$ & $ 69.0 $ & 9.29 & 0.76 & 12.3 & $ -18.8$ & $ 2.3 $ \\ 
$ 13.5$ & $ 69.0 $ & 5.85 & 1.05 & 5.6 & $ -72.8$ & $ 5.2 $ \\ 
$ 1.5$ & $ 69.0 $ & 3.52 & 0.44 & 7.9 & $ -45.8$ & $ 3.6 $ \\ 
$ -22.5$ & $ 69.0 $ & 2.00 & 0.21 & 9.4 & $ -52.7$ & $ 3.1 $ \\
$ -46.5$ & $ 69.0 $ & 1.92 & 0.23 & 8.3 & $ 86.5$ & $ 3.4 $ \\ 
$ 13.5$ & $ 81.0 $ & 5.06 & 0.45 & 11.3 & $ -73.1$ & $ 2.5 $ \\ 
$ 1.5$ & $ 81.0 $ & 2.75 & 0.44 & 6.2 & $ -32.5$ & $ 4.6 $ \\ 
$ -22.5$ & $ 81.0 $ & 2.18 & 0.30 & 7.3 & $ -36.2$ & $ 3.9 $ \\ 
$ 25.5$ & $ 93.0 $ & 4.15 & 0.55 & 7.5 & $ 23.1$ & $ 3.8 $ \\ 
$ 1.5$ & $ 93.0 $ & 2.70 & 0.63 & 4.3 & $ -22.7$ & $ 6.7 $ \\ 
\enddata
\label{p3:n2024table}
\end{deluxetable}

\end{document}